\documentclass[aps,prd,nofootinbib,showkeys,preprint,floatfix]{revtex4}

\usepackage{mathrsfs}

\usepackage{amssymb}
\usepackage{amsmath}
\usepackage{amsfonts}
\usepackage{graphicx}
\usepackage{color}
\usepackage{xspace}
\usepackage{ulem}
\usepackage{lscape}
\usepackage{wasysym}
\usepackage{hyperref}

\usepackage{multirow,rotating}
\usepackage{dsfont}
\usepackage{booktabs}

\allowdisplaybreaks[4]

\def\gsim{\raise0.3ex\hbox{$\;>$\kern-0.75em\raise-1.1ex\hbox{$\sim\;$}}}
\def\lsim{\raise0.3ex\hbox{$\;<$\kern-0.75em\raise-1.1ex\hbox{$\sim\;$}}}

\def\znbb{0\nu\beta\beta}
\def\meff{\langle m_{\nu} \rangle}

\newcommand{\ba}[1]{\begin{eqnarray} \label{(#1)}}
\newcommand{\ea}{\end{eqnarray}}

\newcommand{\AddrAHEP}{
  {\it AHEP Group, Instituto de F\'{\i}sica Corpuscular --
    C.S.I.C./Universitat de Val{\`e}ncia \\
    Edificio de Institutos de Paterna, Apartado 22085,
  E--46071 Val{\`e}ncia, Spain}}

\def\gsim{\raise0.3ex\hbox{$\;>$\kern-0.75em\raise-1.1ex\hbox{$\sim\;$}}}
\def\lsim{\raise0.3ex\hbox{$\;<$\kern-0.75em\raise-1.1ex\hbox{$\sim\;$}}}

\begin{document}

\preprint{IFIC/15-30}

\title{$SU(5)$-inspired double beta decay}

\author{Renato M. Fonseca}\email{renato.fonseca@ific.uv.es}\affiliation{\AddrAHEP}
\author{Martin Hirsch} \email{mahirsch@ific.uv.es}\affiliation{\AddrAHEP}



\begin{abstract}
\vskip3mm The short-range part of the neutrinoless double beta
amplitude is generated via the exchange of exotic particles, such as
charged scalars, leptoquarks and/or diquarks. In order to give a
sizeable contribution to the total decay rate, the masses of these
exotics should be of the order of (at most) a few TeV. Here, we
argue that these exotics could be the ``light'' (i.e weak-scale)
remnants of some $B-L$ violating variants of $SU(5)$. We show that
unification of the standard model gauge couplings, consistent with
proton decay limits, can be achieved in such a setup without the need
to introduce supersymmetry.  Since these non-minimal $SU(5)$-inspired
models violate $B-L$, they generate Majorana neutrino masses and therefore
make it possible to explain neutrino oscillation data. The ``light'' coloured
particles of these models can potentially be observed at the LHC, and
it might be possible to probe the origin of the neutrino masses with
$\Delta L=2$ violating signals. As particular realizations of this idea, 
we present two models, one for each of the two possible tree-level topologies
 of neutrinoless double beta decay.

\end{abstract}

\maketitle


\section{Introduction}

Neutrinoless double beta decay ($\znbb$) is usually considered as a
probe to constrain (or measure) the Majorana neutrino mass scale.
However, a non-zero amplitude for $\znbb$ decay is generated in any
extension of the standard model with lepton number violation (for a
recent review see, for example \cite{Deppisch:2012nb}) and, in
general, the mass mechanism is not necessarily the dominant
contribution to the decay rate. 

The short-range part of the $\znbb$ decay rate \cite{Pas:2000vn} is
generated via the exchange of exotic particles, such as leptoquarks
and/or diquarks plus possibly some exotic fermions.  The ``heavy''
mediators in these diagrams must have masses of at most a few TeV in
order to give a sizeable contribution to the $\znbb$ decay rate, and
thus they can potentially be produced and studied at the LHC
\cite{Helo:2013dla,Helo:2013ika}. A complete list of all (scalar)
mediated short-range contributions to $\znbb$ decay has been given in
\cite{Bonnet:2012kh}. One can understand the results of this work
\cite{Bonnet:2012kh} as a {\it bottom-up} reconstruction of all
possible particle physics models with tree-level $\znbb$ decay. In
this paper we take the opposite approach and study instead the $\znbb$
decay in a {\it top-down} approach; in other words, we explore the
possibility that some of these exotics could be the ``light'' remnants
of multiplets in some unified theory.

Minimal $SU(5)$ accidentally conserves $B-L$
\cite{Mohapatra:1986uf}. Once $SU(5)$ is broken to the standard model
group ($G_{SM}$), baryon number violating processes such as proton
decay occur, but the combination $B-L$ is still conserved. Thus,
neutrinos are as massless in minimal $SU(5)$ as in the standard model
(SM) and remain so also after $SU(5)$ breaking. Adding $SU(5)$
singlets a type-I seesaw mechanism
\cite{Minkowski:1977sc,GellMann:1980vs,Yanagida:1979as,Mohapatra:1979ia}
can be generated.  While certainly theoretically attractive to many,
this scenario leaves as its only prediction that neutrinoless double
beta decay should be observed. \footnote{Albeit without fixing the
  $\znbb$ decay half-life, unless further assumptions about the active
  neutrino mass spectrum are made.}

Allowing for larger $SU(5)$ multiplets, however, $B-L$ is no longer
conserved in $SU(5)$. Perhaps adding a scalar ${\boldsymbol{15}}$ is
the simplest way to obtain a model with $B-L$ violation {\em in the
  $SU(5)$ symmetric phase}.  Indeed, it is possible to write down the
Yukawa interaction
$\boldsymbol{\bar{5}}_{F}\cdot\boldsymbol{15}\cdot\boldsymbol{\bar{5}}_{F}$
as well as a scalar trilinear term
$\mathbf{5}\cdot\boldsymbol{15}^{*}\cdot\mathbf{5}$, with the
$\boldsymbol{\bar{5}}_{F}$ containing the SM $d^{c}$ and $L$ fermion
fields and the $\mathbf{5}$ containing the SM Higgs doublet $H$. In
terms of $G_{SM}$ representations, we have both the
$LLS_{\mathbf{1},\mathbf{3},1}$ and
$HHS_{\mathbf{1},\mathbf{3},1}^{*}$ interactions, where
$S_{\mathbf{1},\mathbf{3},1}\subset\boldsymbol{15}$ \footnote{We use
  $S$ to denote a scalar and $\psi$ for a fermion, subscripts are the
  transformation properties (or charge) under the SM group in the
  order $SU(3)_c \times SU(2)_L \times U(1)_Y$. For $SU(5)$ multiplets
  we add a subscript ``${F}$'' for fermions, no subscript for
  scalars.}, therefore a type-II seesaw
\cite{Schechter:1980gr,Magg:1980ut,Mohapatra:1980yp,Cheng:1980qt} will
be generated, yielding the effective operator $LLHH$ (or
$\boldsymbol{\bar{5}}_{F}\cdot\boldsymbol{\bar{5}}_{F}\cdot\mathbf{5}\cdot\mathbf{5}$
in terms of $SU(5)$ fields). Many other examples of $B-L$ violating
$SU(5)$ models can be constructed using larger representations.

Extrapolating the SM gauge couplings to larger energies with only the
SM particle content fails to achieve unification. Thus, no consistent
model of grand unification can be built without adding new particles
and/or interactions at some so-far unexplored energy scale. One of the
most cited possibility to achieve gauge coupling
unification (GCU) is to extend the SM to the minimal
supersymmetric standard model \cite{Amaldi:1991cn}. However, it has
been known for a long time that also many non-SUSY scenarios can lead
to GCU (for an early reference with new states at TeV, see
\cite{Amaldi:1991zx}; for an early reference with a left-right
symmetric intermediate stage at ($10^{10}-10^{11}$) GeV see
\cite{Brahmachari:1991np}).
For a discussion of GCU in non-SUSY $SU(5)$-based models
see also \cite{Dorsner:2005fq,Dorsner:2006hw}.

In this paper we show that non-minimal $SU(5)$-based extensions of the
SM can lead to good GCU, provided some beyond-SM (coloured) multiplets
are accidentally light. The very same TeV-scale remnants of these
non-minimal models, responsible for GCU, could be (some of) the
mediators of the short-range double beta decay amplitude, leading to
lepton number violation (LNV) at the electro-weak scale. We call this
``$SU(5)$-inspired'' double beta decay. Different models realizing this
idea can be constructed. Since short-range diagrams for $\znbb$ decay
always fall into one of only two possible tree-level topologies
\cite{Bonnet:2012kh}, we will discuss two particular variants, namely
one model for topology-I (T-I) and one for topology-II (T-II).

The main motivation to study these $SU(5)$-inspired models is
that they are experimentally falsifiable at the LHC and, possibly, in
upcoming lepton flavour violation searches in the following sense.
First, current limits on the half-lives of $\znbb$ decay in $^{76}$Ge
\cite{Agostini:2013mzu} and $^{136}$Xe
\cite{Albert:2014awa,Gando:2012zm,Asakura:2014lma} are of the order of
$(1-2)\times 10^{25}$ years, resulting in lower limits on the effective
mass scale of the underlying operator in the range of $M_{\rm eff}
\simeq (2-2.5) g_{\rm eff}^{4/5}$ TeV for our two example models,
while an observation with a half-life below roughly $10^{27}$ years
implies $M_{\rm eff} \lsim (3-3.8) g_{\rm eff}^{4/5}$ TeV.
Here, $g_{\rm eff}$ is the mean of the couplings entering in the
$\znbb$ decay diagram(s), see below, with $g_{\rm eff}$ required to be roughly of the
order ${\cal O}(0.1-1)$ to give an observable decay rate.  Negative
searches at the LHC in run-II will allow the covering of this range of masses
\cite{Helo:2013dla,Helo:2013ika}. Second, oscillation experiments have
shown\footnote{See for example the recent fit \cite{Forero:2014bxa}.}
that lepton flavour is violated (LFV). If our models are to explain
neutrino data consistently, LFV violating entries in the Yukawa
couplings are therefore required. These will lead to non-zero rates in
processes such as $\mu \to e \gamma$, with expected rates that could
be of the order of the current experimental limit \cite{Adam:2013mnn}.

We should also add a disclaimer about naturalness here. Standard
$SU(5)$ suffers from what is known as the doublet-triplet splitting
problem, i.e. the fact that the SM Higgs has a mass of $m_h \simeq
125$ GeV \cite{Aad:2012tfa,Chatrchyan:2012ufa} while the coloured
triplet in the ${\bf 5}$ must have a GUT scale mass. Although several
solutions to this problem have been suggested (for a short review see
\cite{Randall:1995sh}), we do not concern ourselves with any
particular one. Instead, we view this simply as a fine-tuning problem
and, in fact, to make the exotic particles in our models light
will require, in general, additional fine-tunings. We assured
ourselves, however, that this can be done consistently for all the
light states (details can be found in appendix
\hyperref[sec:AppendixB]{B}).

The rest of this paper is organized as follows. In section
\ref{sec:MdlA} we discuss our model based on topology-I.  It has three
new multiplets when compared to minimal $SU(5)$, one of which could
contain a good candidate for the cold dark matter in the Universe. The
model also has a coloured octet fermion and a scalar leptoquark at the
TeV scale, both of which can be produced at the LHC. Neutrino masses
are dominated by 2-loop diagrams. Section \ref{sec:MdlB} discusses the
model based on T-II. More light states exist in this variant model,
leading to a more diverse phenomenology.  As in model T-I, in model
T-II neutrino masses are generated at 2-loop order.  There are new
coloured sextets, which have particularly large LHC cross sections and
there is also a doubly charged scalar. We then close the paper with a
short conclusion.  Many of the technical aspects of our work are
deferred to the appendices \hyperref[sec:AppendixA]{A},
\hyperref[sec:AppendixB]{B} and \hyperref[sec:AppendixC]{C}, where we
give details of the Lagrangians of the two models, discuss briefly
some aspects of fine-tuning in $SU(5)$ and present a table with the
decompositions of larger $SU(5)$ multiplets.

\section{A simple model with a scalar quadruplet}
\label{sec:MdlA}

We start this section with some preliminary comments.  Minimal $SU(5)$
puts the three standard model families of fermions into three copies
of ${\boldsymbol{\bar 5}_F}$ and ${\boldsymbol{10}_F}$. To generate
the SM Dirac masses at least one scalar ${\bf 5}$ is needed.  Although
not strictly speaking ``minimal'' we allow, in principle, also for the
presence of a scalar ${\bf 45}$ at the GUT scale. This ${\bf 45}$ is
added for the sole purpose of generating a Georgi-Jarlskog factor for
fermion masses \cite{Georgi:1979df} --- see appendix
\hyperref[sec:AppendixB]{B}.  In addition, at least one scalar ${\bf
  24}$ is needed to break $SU(5)$ to the SM group. As mentioned
already in the introduction, the Lagrangian of this minimal model
conserves $B-L$ \cite{Mohapatra:1986uf}.

Both our models need to introduce larger representations. For
completeness we give in the appendix \hyperref[sec:AppendixC]{C} the
decomposition into SM group representations of all $SU(5)$ multiplets
up to the ${\bf 75}$. Once these larger multiplets are added, $B-L$
will be violated, as discussed with the example of the ${\bf 15}$ in
the introduction.

\begin{figure}[t]
\begin{center}
\includegraphics[scale=0.95]{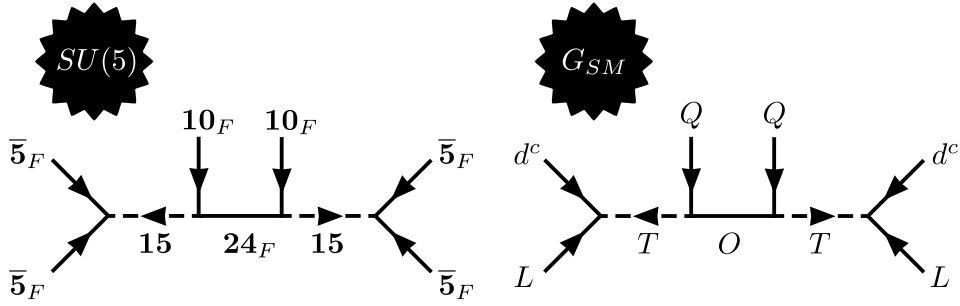}
\end{center}
\caption{Double beta decay short-range diagram in $SU(5)$ language
  (left) and in the $G_{SM}$ language (right). The fields
  ${\boldsymbol{\bar 5}_F}$ and ${\boldsymbol{10}_F}$ contain the
  standard model fermions, while
  ${\boldsymbol{T}}=S_{\mathbf{3},\mathbf{2},1/6}$ and
  ${\boldsymbol{O}}=\psi_{\mathbf{8},\mathbf{1},0}$ are the ``light''
  pieces coming from the ${\bf 15}$ and ${\boldsymbol{24}_F}$,
  respectively.}
\label{fig:MdlA}
\end{figure}

Let us now turn to our first example model, which we call T-I in the
following. To generate a topology I $\znbb$ decay, an exotic fermion
is needed. We thus add to minimal $SU(5)$ three multiplets: two
scalars (${\bf 15}$ and ${\bf 70}$) and one fermion
($\boldsymbol{24}_F$).  The ${\bf 15}$ and ${\boldsymbol{24}_F}$ will
generate $\znbb$ decay and neutrino masses, while the ${\bf 70}$ can
play the role of a dark matter candidate, as explained later on.
Details of the Lagrangian of the model are given in appendix
\hyperref[sec:AppendixB]{B}.  Here we only discuss the most relevant
terms.  Consider first the interaction among the ${\bf 15}$, the
${\boldsymbol{24}_F}$ and the SM fermions:
\begin{eqnarray}\label{eq:Lag5MA}
{\cal L} & = & 
{\widehat y}_{ij}^{(5)} {\bf\bar 5}_{{F},i}{\boldsymbol{\bar 5}}_{{F},j}{\bf 15}
+  {\widehat y}_{i}^{(7)}{\bf 10}_{{F},i}{\boldsymbol{24}_F}{\bf 15^*}  + \cdots \, .
\end{eqnarray}
If we add to eq. (\ref{eq:Lag5MA}) a Majorana mass term for 
the ${\boldsymbol{24}_F}$,
\begin{equation}\label{eq:m24}
{\cal L}^{M} = m_{\bf 24} {\boldsymbol{24}_F}{\boldsymbol{24}_F}\, , 
\end{equation} 
an effective $d=9$ operator is generated, which in $SU(5)$ language reads
\begin{equation}\label{eq:O9SU5MdA}
{\cal O}_9 \propto ({\boldsymbol{\bar 5}_F}{\boldsymbol{\bar 5}_F})({\boldsymbol{10}_F})({\boldsymbol{10}_F})
({\boldsymbol{\bar 5}_F}{\boldsymbol{\bar 5}_F})
\end{equation}
--- see the left-hand side of fig. (\ref{fig:MdlA}).  Under the SM
group, the ${\bf 15}$ and the $\boldsymbol{24}_F$ break as
$S_{\mathbf{3},\mathbf{2},1/6} + \cdots \equiv {\boldsymbol{T}} +
\cdots$ and $\psi_{\mathbf{8},\mathbf{1},0}+ \cdots \equiv
      {\boldsymbol{O}} + \cdots$, respectively. The above Lagrangian
      equations (\ref{eq:Lag5MA}) and (\ref{eq:m24}) then contain the
      following terms:
\begin{eqnarray}\label{eq:LagSMMdlA}
{\cal L} & = & 
  y_{ij}^{(4)}
 L_i d^c_j \boldsymbol{T}
 + 
 y_{i}^{(5)} 
 Q_i \boldsymbol{O} \boldsymbol{T}^*
+ m_{O}  {\boldsymbol{O}} {\boldsymbol{O}}\, ,
\end{eqnarray}
with $y^{(4)}={\widehat y}^{(5)}$, $y^{(5)}={\widehat y}^{(7)}$ and
$m_{O}=m_{\bf 24}$ in the $SU(5)$ symmetric phase.

Eq. (\ref{eq:LagSMMdlA}) will produce an operator that generates a
contribution to double beta decay via
\begin{equation}\label{eq:O9SU5SMMdA}
{\cal O}^{4-i}_{11} \propto (L d^c)(Q)(Q)(L d^c)\, .
\end{equation}
Here, the subscript ``11'' indicates the number of this $\Delta L=2$
operator in the list defined by Babu \& Leung \cite{Babu:2001ex},
while the superscript ``4-i'' identifies the double beta decay
decomposition according to the list of \cite{Bonnet:2012kh}.  The
diagram which generates this operator is shown in
fig. (\ref{fig:MdlA}), on the right. In order for this diagram to give
a sizeable contribution to the total $\znbb$ decay amplitude,
${\boldsymbol{T}}$ and ${\boldsymbol{O}}$ must have masses of the TeV
order. The amplitude for this diagram has been calculated in
\cite{Bonnet:2012kh}: the limit from $^{136}$Xe
\cite{Albert:2014awa,Gando:2012zm,Asakura:2014lma} results in $M_{\rm
  eff} \gsim 2.5 g_{\rm eff}^{4/5}$ TeV, where $M_{\rm eff} = (m_T^4
m_O)^{1/5}$ and $g_{\rm eff}$ is the (geometric) mean of the four
couplings entering the diagram. In addition to this diagram, $\znbb$
decay will receive also a contribution from the mass mechanism.  We
will compare the short-range contribution and mass mechanism below,
when discussing neutrino masses in this model.

We now turn to the discussion of the ${\bf 70}$ scalar
representation. The standard model does not contain a particle
candidate for the cold dark matter (DM). It is, however, quite
straightforward to identify the basic requirements for SM multiplets
to contain viable dark matter candidates. As shown in table
\ref{tab:decopm} in the appendices, the ${\bf 70}$ is one of the
smallest $SU(5)$ multiplet containing an $SU(2)_L$ quadruplet, ${\bf
  70} = S_{\mathbf{1},\mathbf{4},1/2} +\cdots \equiv {\boldsymbol{K}}
+ \cdots$. $S_{\mathbf{1},\mathbf{4},1/2}$ contains one electrically
neutral state, ${\boldsymbol{K}^0}$, which after the breaking of
$SU(2)_L$ can play the role of the cold dark matter.

Of course, only particles stable over cosmologically long times can be
dark matter. In the $SU(5)$ phase, model T-I allows, in principle, a
quartic scalar term ${\bf 5}\cdot {\bf 5}\cdot{\bf 5^*}\cdot {\bf
  70^*}$ which could induce the decay ${\boldsymbol{K}^0} \to h h
h$. To eliminate this coupling, and all other couplings linear or
cubic in ${\bf 70}$, one has to postulate a $Z_2$ symmetry. We simply
assume ${\boldsymbol{K}}$ to be odd under this $Z_2$,
\footnote{A non-zero vacuum expectation value (VEV) of the
  ${\boldsymbol{K}^0}$ would break this $Z_2$ spontaneously and, thus,
  has to be avoided. For $m_K^2 > 0$ one expects a zero VEV to be the
  preferred solution of the tadpole equations.}  while all other
particles of the model are even. We will not discuss further details
of the phenomenology of ${\boldsymbol{K}^0}$ as a DM candidate; these
have been worked out in \cite{Cirelli:2005uq}. Note that according to
\cite{Cirelli:2005uq}, the neutral member of ${\boldsymbol{K}}$ can
play the role of DM if $m_{K^0} \simeq 2.4$ TeV.

Adding complete $SU(5)$ multiplets to the SM does not change GCU,
since the $\beta$ coefficients of all three gauge couplings change by
the same amount. However, after $SU(5)$ breaking, the masses of the
different $G_{SM}$ multiplets contained within each $SU(5)$
representation may be different, yielding a large number of
possibilities to achieve GCU.

\begin{figure}[t]
\includegraphics[scale=0.70]{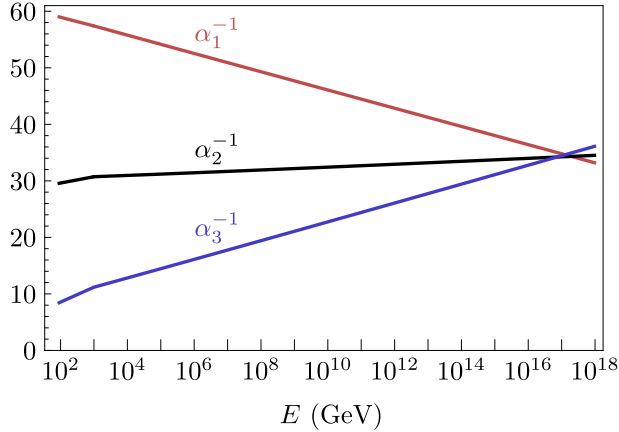}
\caption{Gauge coupling unification in model T-I. In this
  calculation the new states are assumed to have masses around $M \sim
  {\cal O}(1)$ TeV. The running includes 2-loop $\beta$ coefficients,
  but does not consider any GUT scale thresholds.}
\label{fig:gcuA}
\end{figure}

The $\beta$ coefficients for the running of the gauge couplings, 
\footnote{In general, ignoring the effect of Yukawa interactions, the
  running gauge couplings $g_{i}$ change with the logarithm of the
  energy scale $t\equiv\log E$ as follows:
  $dg_{i}/dt=b_{i}g_{i}^{3}/\left(4\pi\right)^{2}
  +b_{ij}g_{i}^{3}g_{j}^{2}/\left(4\pi\right)^{4}$.}  including the
contributions from ${\boldsymbol{T}}$, ${\boldsymbol{O}}$ and
${\boldsymbol{K}}$ are
\begin{eqnarray}
\label{biMA}
b_i=
\begin{pmatrix}
\frac{13}{3} \\  -1 \\ \frac{-14}{3} 
\end{pmatrix}
& , \qquad &
b_{ij} =
\begin{pmatrix}
 \frac{326}{75}  & 12  & \frac{28}{3} \\
 4 & 94 & 20 \\
 \frac{7}{6} & \frac{15}{2} & \frac{88}{3} 
\end{pmatrix}\, ,
\end{eqnarray}
at 1-loop and 2-loop order, respectively.  Fig. (\ref{fig:gcuA}) shows
the resulting running of the inverse gauge couplings as a function of
energy, assuming that the masses of ${\boldsymbol{T}}$,
${\boldsymbol{O}}$ and ${\boldsymbol{K}}$ are of the TeV order. Here,
we neglect the small 2-loop contributions from Yukawa couplings for
simplicity, and we also do not consider possible corrections from
GUT-scale thresholds.

As fig. (\ref{fig:gcuA}) shows, the model nicely unifies with an
estimated GUT scale of $m_G \simeq 10^{17}$ GeV, which gives an
estimated half-life for gauge mediated proton decay of $T_{1/2}(p) \sim
10^{38}$ years, albeit with a large uncertainty. Recall that the current
constraint from Super-Kamiokande is $\tau_{p\to\pi^{0}e^{+}}\gtrsim
10^{34}$ years~\cite{Abe:2013lua}.

\begin{figure}[t]
\includegraphics[scale=0.95]{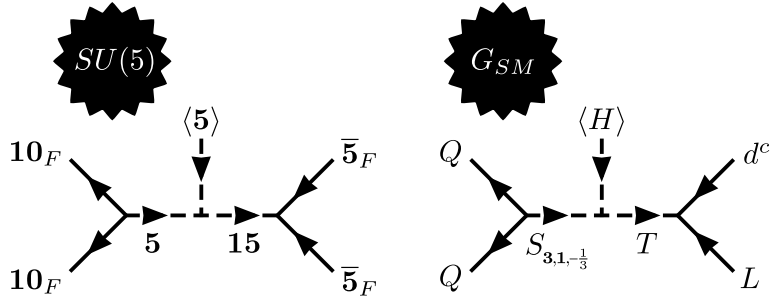}
\caption{Scalar-mediated proton decay in model T-I. Mixing induced via
  the SM Higgs VEV between ${\bf S_{\mathbf{1},\mathbf{3},-1/3}}$ and
  ${\boldsymbol{T}}$ can generate an effective diquark coupling for
  ${\boldsymbol{T}}$.}
\label{fig:pdecA}
\end{figure}

However, while the model is automatically safe from the gauge mediated
proton decay diagrams, we have to impose one more constraint on the
model, due to scalar mediated proton decay.  The triple-scalar term
${\bf 5}\cdot {\bf 5}\cdot {\bf 15^*}$, see fig. (\ref{fig:pdecA}),
leads to a vertex ${S_{\mathbf{3},\mathbf{1},-1/3}} H
{\boldsymbol{T}}$. After electro-weak symmetry breaking this can be
considered as an effective mixing of ${\boldsymbol{T}}$ with the
coloured triplet ${S_{\mathbf{3},\mathbf{1},-1/3}}$. Note, however,
that in the full model there is more than one contribution to this
vertex, since the field content of the model also allows the writing
of the interaction ${\bf 5}\cdot{\bf 5}\cdot {\bf 15^*}\cdot{\bf 24}$.
Furthermore, there is a second ${S_{\mathbf{3},\mathbf{1},-1/3}}$ inside the $\bf {45}$
which needs to be taken into consideration. As such, if we call $\mu_1$ and  $\mu_2$ the effective
interactions of these two heavy colored scalars (assumed to have a degenerate mass of the order of the GUT scale $m_G$) with  $H$ and ${\boldsymbol{T}}$, then
\begin{align}
\mu_{1} & =2\cos\alpha\,\widehat{h}_{\mathbf{5}\cdot\mathbf{5}\cdot\mathbf{15}^{*}}+2\cos\alpha\,\widehat{\lambda}_{\mathbf{5}\cdot\mathbf{5}\cdot\mathbf{15}^{*}\cdot\mathbf{24}}\left\langle \mathbf{24}\right\rangle -\sin\alpha\sum_{a=1,2}\widehat{\lambda}_{\left[\mathbf{5}\cdot\mathbf{45}\cdot\mathbf{15}^{*}\cdot\mathbf{24}\right]_{a}}\left\langle \mathbf{24}\right\rangle \\
\mu_{2} & =\cos\alpha\sum_{a=1,2}\widehat{\lambda}_{\left[\mathbf{5}\cdot\mathbf{45}\cdot\mathbf{15}^{*}\cdot\mathbf{24}\right]_{a}}\left\langle \mathbf{24}\right\rangle -2\sin\alpha\left(\widehat{h}_{\mathbf{45}\cdot\mathbf{45}\cdot\mathbf{15}^{*}}+\sum_{a=1,2}\widehat{\lambda}_{\left[\mathbf{45}\cdot\mathbf{45}\cdot\mathbf{15}^{*}\cdot\mathbf{24}\right]_{a}}\left\langle \mathbf{24}\right\rangle \right)
\end{align}
where $\alpha$ is the angle controlling the admixture of the ${S_{\mathbf{1},\mathbf{2},1/2}}$ representations  in the  $\bf {5}$ and  $\bf {45}$ forming the light state $H$, $\widehat{h}$ ($\widehat{\lambda}$) represent the trilinear (quartic) scalar couplings of the $SU(5)$ fields indicated in subscript, and the index $a$ keeps track of the different gauge invariant contractions of the representations. The VEV of the $\mathbf{24}$ must be in the SM-singlet direction.

Since the ${S_{\mathbf{3},\mathbf{1},-1/3}}$ fields has diquark
couplings, a proton decay diagram is induced by this
${S_{\mathbf{3},\mathbf{1},-1/3}} \leftrightarrow {\boldsymbol{T}}$ mixing. Consistency with proton decay limits can be converted into an
upper limit on the sum of the two $\mu_i$ couplings, $\mu\equiv\mu_{1}+\mu_{2}$, which is, very roughly, of the order of
\begin{equation}\label{eq:limh1}
\frac{\mu}{m_{G}}\lsim2.5\times10^{-6}\left(\frac{0.1}{y_{11}^{(4)}}\right)\left(\frac{2\hskip0.5mm{\rm TeV}}{m_{T}}\right)^{2}\, .
\end{equation}
Here we assume that the typical value of $y_{11}^{(4)}$ is 
${\cal O}(0.1)$. Much smaller values, say $y_{11}^{(4)} \simeq y_e$, 
would avoid the fine-tuning of eq. (\ref{eq:limh1}), but at the 
same time they would render the $\znbb$ diagram in fig. (\ref{fig:MdlA}) 
unobservable. 

We note that a very similar discussion about scalar-mediated proton
decay can be found in \cite{Dorsner:2005fq,Babu:2012vc}. Reference \cite{Dorsner:2005fq} considers also $SU(5)$, 
while \cite{Babu:2012vc} discusses an $SO(10)$ based model. In \cite{Babu:2012vc}, however, the authors argue against
the existence of light colour-triplet scalars on the basis of this constraint, while we accept eq. (\ref{eq:limh1}) as just one
(more) fine-tuning.

Let us now discuss neutrino masses. The ${\bf 15}$ also contains the
$S_{\mathbf{1},\mathbf{3},1}$, which is often denoted as $\Delta$ in
neutrino physics \cite{Mohapatra:1998rq,Valle:2015pba}. Thus, after
integrating out the heavy $\Delta$, effectively a Weinberg operator
\cite{Weinberg:1979sa} is generated. After the breaking of $SU(2)_L$,
one thus finds a type-II seesaw contribution to the neutrino mass. One
can estimate this contribution to be of the order of
\begin{equation}\label{eq:mnuTII}
(m_{\nu})_{ij} \simeq {\widehat y}_{ij}^{(5)} v_{SM}^2
\frac{\mu_{\Delta}}{m_{\Delta}^2} 
\sim 0.3 \hskip1mm {\rm meV} \, . 
\end{equation}
For ${\widehat y}_{ij}^{(5)}=1$ and $\mu_{\Delta}=m_{\Delta}\simeq m_G$,
with a GUT scale of, see fig. (\ref{fig:gcuA}), $m_G \simeq 10^{17}$
GeV. This is too small to explain solar neutrino oscillations by a
factor of roughly 30. 

Note that the effective $\mu_{\Delta}$ contains contributions from the
same couplings (${\bf 5}\cdot{\bf 5}\cdot {\bf 15^*}$, etc.) as
discussed just above for the scalar induced proton decay. The
different $SU(5)$ terms contribute, however, with different
Clebsch-Gordon coefficients to $\mu_{\Delta}$ than those appearing in
$\mu$, see again appendix \hyperref[sec:AppendixB]{B} for
details. Thus, $\mu$ and $\mu_{\Delta}$ can easily take very different
values. Of course, if $\mu_{\Delta}$ were to obey a limit similar to
eq. (\ref{eq:limh1}), contributions to neutrino masses would actually
be much smaller than the numerical value in eq. (\ref{eq:mnuTII}).

\begin{figure}[t]
\begin{center}
\includegraphics[scale=0.95]{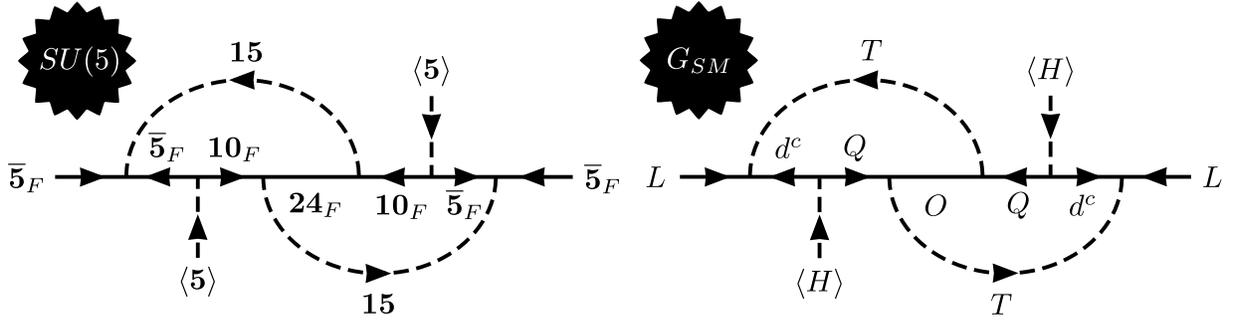}
\end{center}
\caption{2-loop neutrino mass diagram in model T-I. To the left, the 
$SU(5)$ origin of the diagram; to the right the contributions to $m_{\nu}$ 
from the light states of the model.}
\label{fig:MdlAmnu}
\end{figure}

The failure to have a sufficiently large neutrino mass to explain 
solar/atmospheric data via the type-II seesaw contribution
\footnote{Unless $m_{\Delta}$ is tuned to be much smaller than the GUT
  scale, which would harm the successful GCU.} does not imply that the
model cannot explain neutrino oscillation data. This is due to the
two-loop contribution to the neutrino mass matrix, involving the
leptoquark and the coloured fermion, shown in fig. (\ref{fig:MdlAmnu})
(on the right). The left diagram shows the $SU(5)$ origin of this
diagram, the diagram to the right is in the SM phase.

Using the general analysis of 2-loop diagrams of \cite{Sierra:2014rxa}
one can calculate the contribution of this diagram to the neutrino
mass matrix. However, since a 2-loop model based on this diagram has
been discussed recently in \cite{Angel:2013hla}, we will not repeat
all the details here and show only a few rough estimates. For 
$m_T \simeq m_O \gg m_b$ fig. (\ref{fig:MdlAmnu}) gives approximately 
\begin{equation}\label{eq:estmnu}
(m_{\nu})_{ij} \simeq \frac{N_c}{(16 \pi^2)^2}\frac{1}{\Lambda_{\rm LNV}} 
\Big( m_{d_k}m_{d_m} y_{i k}^{(4)}y_{j m}^{(4)}y_{k}^{(5)}y_{m}^{(5)} 
+ (i  \leftrightarrow j ) \Big).
\end{equation}
$N_c$ is a colour factor ($N_c=3$ for this diagram) and logarithmic
terms from the loop integrals have been neglected for
simplicity. $\Lambda_{\rm LNV}$ is the mean of the masses of $m_T
\simeq m_O$.  Note that eq. (\ref{eq:estmnu}) produces three non-zero
neutrino masses, which in leading order will be proportional to
$m_{\nu_{1,2,3}} \propto m_bm_d , m_b m_{s}, m_{b}^2$. While it is possible 
to fit quasi-degenerate neutrino masses with eq. (\ref{eq:estmnu}), 
the hierarchy in down quark masses leads us to expect that neutrino 
masses follow a normal hierarchical neutrino spectrum in model T-I. For
$\Lambda_{\rm LNV} \simeq 1$ TeV, $|y_{3}^{(4)}| \sim y_{3}^{(5)}
\simeq 0.06$ and $|y_{2}^{(4)}| \sim y_{2}^{(5)} \simeq 0.26$, where
$|y_{k}^{(4)}|=\sqrt{\sum_j (y_{jk}^{(4)})^2}$, one obtains
$m_{\nu_{2,3}} \simeq (9\times 10^{-3},5\times 10^{-2})$ eV, correctly
reproducing the atmospheric and solar neutrino mass scales (for normal
hierarchy).

With two Yukawa vectors ($y_{i 3}^{(4)}$ and
$y_{i 2}^{(4)}$) contributing (dominantly) to the flavour
structure of $(m_{\nu})_{ij}$, there are four independent
ratios, which we arbitrarily choose to be
$(y_{13}^{(4)}/y_{33}^{(4)},y_{23}^{(4)}/y_{33}^{(4)})$ and
$(y_{12}^{(4)}/y_{32}^{(4)},y_{22}^{(4)}/y_{32}^{(4)} )$. With these
we can easily fit the observed neutrino angles.  Many solutions exist
and can be found in a simple numerical scan. We only quote one 
particular example as a proof-of-principle. Choosing the absolute
values of the Yukawa couplings as mentioned above,
$(y_{13}^{(4)}/y_{33}^{(4)},y_{23}^{(4)}/y_{33}^{(4)}) \sim
(0.03,-1.7)$ and
$(y_{12}^{(4)}/y_{32}^{(4)},y_{22}^{(4)}/y_{32}^{(4)} ) \sim (1,1)$
gives all three measured neutrino angles near their best fit points
\cite{Forero:2014bxa}. Nevertheless, we stress again that these numerical
examples give only a rough estimate.

One should also compare the size of the relative contributions to the
$\znbb$ decay rate of the short-range diagram (see
fig. (\ref{fig:MdlA})) with the one of the neutrino mass mechanism
\cite{Helo:2015fba}. The current limits on $\znbb$ decay from
$^{76}$Ge and $^{136}$Xe correspond to an upper limit on the effective
neutrino mass, $\meff$, of roughly $\meff \lsim (0.2-0.4)$ eV,
depending on the choice of nuclear matrix elements
\cite{Muto:1989cd,Faessler:2012ku,Menendez:2011zza}. Experiments on
these nuclei with a sensitivity of $10^{27}$ years would probe $\meff
\simeq 50$ meV.  For a normal hierarchical neutrino mass spectrum, as
in our example fit discussed above, much larger half-lives are
expected. The short range diagram gives a half-life of roughly
$10^{27}$ years for $M_{\rm eff} \gsim 4 g_{\rm eff}^{4/5}$ TeV,
corresponding to $M_{\rm eff} \gsim 1.7$ TeV for $g_{\rm eff}=
\sqrt{y_{11}^{(4)} y_{1}^{(5)}} \simeq 0.35$. The latter can be
completely covered by LHC searches for the signal $l^+l^++4j$, as
discussed below.

We have also estimated Br($\mu\to e\gamma$) for this model. Using the
formulas \cite{Angel:2013hla,Helo:2015fba} we can roughly estimate
Br($\mu\to e\gamma$)$\simeq 3\times 10^{-13}
|y_{13}^{(4)}y_{23}^{(4)*}|^2$ for a leptoquark mass of $M_O \simeq
1$ TeV. This is roughly of the order of the limit expected for the
second phase of the MEG experiment \cite{Adam:2013mnn}. A much more 
detailed discussion about LFV can be found in \cite{Angel:2013hla}.

We now discuss briefly LHC searches for model T-I.  Leptoquarks have
been searched at the LHC in run-I by both the ATLAS
\cite{Stupak:2012aj,ATLAS:2012aq} and the CMS
\cite{CMS:2014qpa,Khachatryan:2015bsa,Khachatryan:2014ura}
collaborations. No positive observation of any leptoquark state has
been reported, apart from a possible 2.6$\sigma$ excess near $m_{\rm
  LQ} \simeq 650$ GeV found in \cite{CMS:2014qpa}. Lower limits on LQ
masses depend on the lepton generation they couple to, and on the
branching ratios of the LQs. Lower limits are roughly $650-1000$ GeV
\cite{Stupak:2012aj,ATLAS:2012aq,CMS:2014qpa,%
  Khachatryan:2015bsa,Khachatryan:2014ura} for searches based on some
pair-produced LQs, depending on assumptions. We also mention the
recent CMS search for singly produced LQ states.  This search excludes
single production of first generation LQs with masses below 1730 GeV
(895 GeV) for leptoquark couplings equal to $y = 1$ ($y=0.4$)
\cite{CMS:2015kda}. Run-2 will improve these numbers in the very near
future.

The coloured fermionic octet, ${\boldsymbol{O}}$, can be pair-produced
at the LHC in gluon-gluon fusion. We have implemented the model in
SARAH \cite{Staub:2012pb,Staub:2013tta} and used the Toolbox
environment \cite{Staub:2011dp} to generate SPheno
\cite{Porod:2003um,Porod:2011nf} and MadGraph files.  We then used
MadGraph5 \cite{Alwall:2014hca} for a quick estimation of the cross
section for pair producing ${\boldsymbol{O}}$ at $\sqrt{s}=8$ and
$\sqrt{s}=13$ TeV.  The kinematics of this decay depends on the mass
hierarchy between ${\boldsymbol{O}}$ and our leptoquark
${\boldsymbol{T}}$, but the decays of ${\boldsymbol{O}}$ will always
lead to a final state containing two hard jets plus a charged lepton
or neutrino (i.e. missing $E_T$).  In total, the signal will thus
consist of four hard jets with $2,1$ or $0$ charged leptons. Due to
the Majorana nature of ${\boldsymbol{O}}$, the ratio $R$ of the number
of same-sign to opposite sign charged lepton events is expected to be
$R=1$.

Also, we expect that the signal will contain lepton flavour violating 
events, i.e final states $e^+\mu^+$ plus jets (and also events with taus). 
This is caused by the flavour structure of the leptoquark couplings 
$y_{i k}^{(4)}$, which is needed to explain neutrino angles and which 
enter in the decay rate of the state ${\boldsymbol{O}}$. 

While there is no dedicated search for this signal at neither CMS nor ATLAS,
we can use the results of the CMS search for right-handed
neutrinos in left-right symmetric models \cite{Khachatryan:2014dka} to
estimate roughly the current sensitivity of LHC to ${\boldsymbol{O}}$.  The CMS
collaboration uses the final state $2l+2j$ to extract lower limits on
the masses of $W_R$ and $\nu_R$, but it mentions explicitly that in the
final state they allow for any number of jets larger or equal to two. 
\footnote{There is a $2.8$$\sigma$ excess in the data at
  $m_{eejj}\simeq 2.2$ TeV, but CMS concludes that it is not in
  agreement with expectations for a left-right symmetric model. Our
  model T-I cannot explain this excess either.}  Below $m_{lljj} \sim
2$ TeV the limits derived by CMS are background dominated. Since one
can expect that backgrounds are smaller for signals with larger number
of jets, we can thus use the derived upper limits on $\sigma(pp\to ll
jj)$ to conservatively estimate lower limits on $m_{
  O}$. Unfortunately, CMS decided to show limits only for $m_{lljj}
\ge 1$ TeV, with upper limits on $\sigma(pp\to ee jj)$ and
$\sigma(pp\to \mu\mu jj)$ both of the order of $\sim 9$ fb for this
lowest mass (summed over both lepton charges).  Comparing with their
fig. (2) \cite{Khachatryan:2014dka}, one can estimate that in the bins
[0.8,1] ([0.6,0.8]) TeV upper limits should be roughly $14$ ($23$)
fb. Comparing with the calculated $\sigma(pp\to
{\boldsymbol{O}}{\boldsymbol{O} })$ of 24 (64) fb for $m_{O}=900$
($800$) GeV and assuming a branching ratio
$\textrm{Br}({\boldsymbol{O}}{\boldsymbol{O}}\to ll+4j) \simeq 25$ \%,
we estimate that the current lower limit on $m_{O}$ should be roughly
in the $800-900$ GeV ballpark. More exact numbers would require a
dedicated search by the experimental collaborations. However, for
$\sqrt{s}=13$ TeV, cross sections will be much larger and we estimate
a dedicated search could be sensitive up to $m_{O} \simeq (2.1-2.2)$
TeV with ${\cal L} \simeq 300$ fb$^{-1}$, corresponding to (16-10)
events before cuts.

\section{Model T-II: Adding coloured sextets}
\label{sec:MdlB}

\begin{figure}[t]
\begin{center}
\includegraphics[scale=0.95]{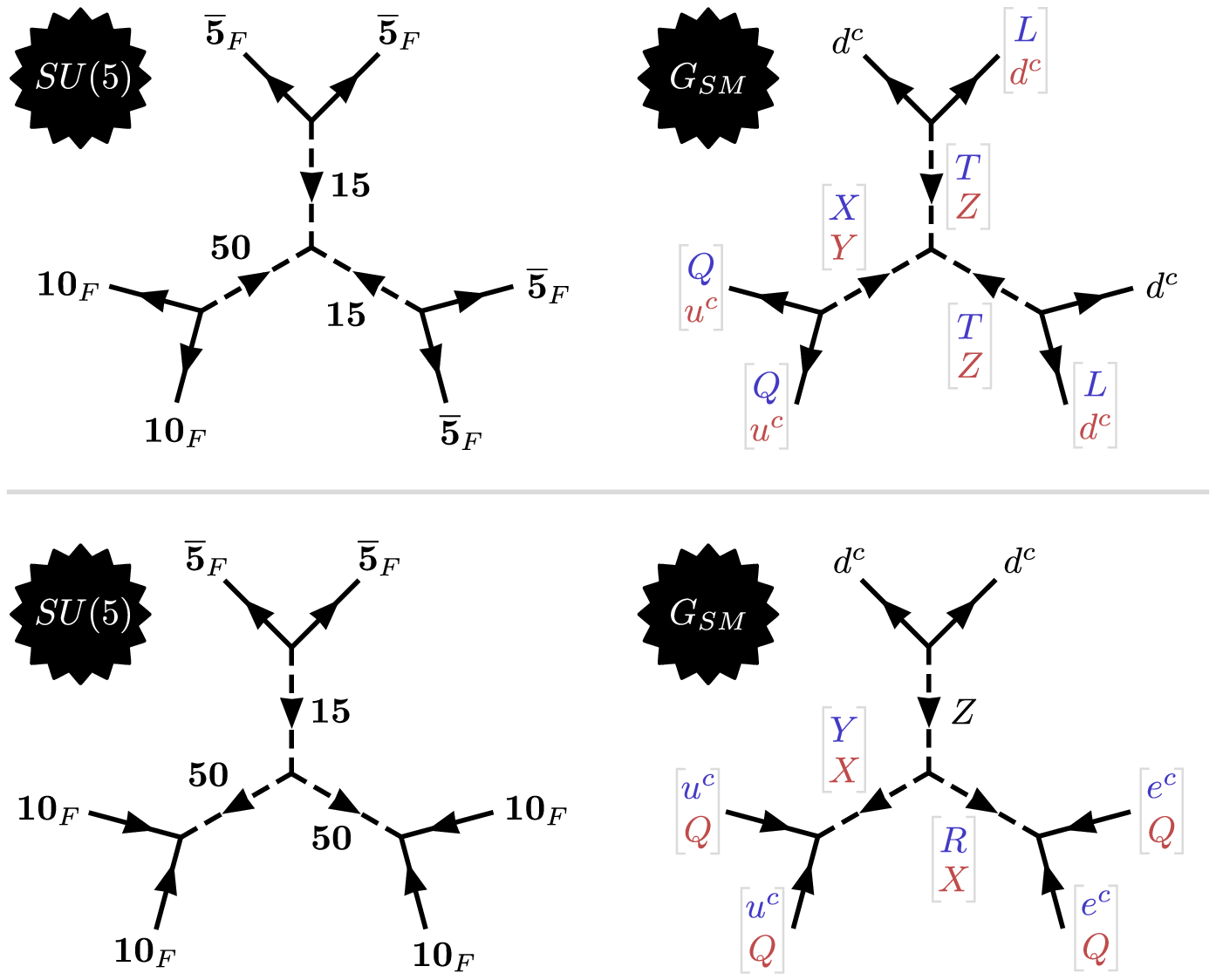}
\end{center}
\caption{Dimension-9 operators in model T-II built with $SU(5)$
  representations (left), and with the corresponding light $G_{SM}$
  pieces (right). Here, in addition to ${\boldsymbol{T}}\equiv
  S_{\mathbf{3},\mathbf{2},1/6}$, we are using the shorthand notation
  ${\boldsymbol{R}}=S_{\mathbf{1},\mathbf{1},-2}$,
  ${\boldsymbol{X}}=S_{\mathbf{6},\mathbf{3},-1/3}$,
  ${\boldsymbol{Y}}=S_{{\bar {\mathbf{6}}},1,4/3}$ and
  ${\boldsymbol{Z}}=S_{{\bar {\mathbf{6}}},1,-2/3}$. These operators
  generate $0\nu\beta\beta$ decay (with the blue/top fields on the
  right diagrams) as well as $n-\overline{n}$ oscillations (with the
  red/bottom fields) --- see text for a discussion.}
\label{fig:MdlCbb}
\end{figure}

Topology-II for $\znbb$ decay consists of diagrams in which 
three scalars are interchanged between the six SM fermions 
that make up the $\znbb$ decay operator \cite{Bonnet:2012kh}. 
For model T-II we therefore add two copies of the scalar ${\bf 15}$ 
plus one scalar ${\bf 50}$ to minimal $SU(5)$. The 
Lagrangian of the model is given in the appendix. Here we 
discuss only the most relevant terms. In the $SU(5)$ phase 
the Lagrangian includes
\begin{eqnarray}\label{LagC}
{\cal L}_{FFS} &=& {\widehat{y}}_{ijk}^{(5)}
          {\boldsymbol{\bar 5}}_{{F},i}{\boldsymbol{\bar 5}}_{{F},j} {\bf 15}_k
   + {\widehat y}_{ij}^{(6)} {\bf 10}_{{F},i}{\bf 10}_{{F},j}{\bf 50} + \cdots \, , \\
{\cal L}_{SSS} &=& {\widehat h}_{ij} ^{(2)}{\bf 15}_i{\bf 15}_j{\bf 50} 
 +  {\widehat h}_{i} ^{(3)} {\bf 15}_i{\bf 50^*} {\bf 50^*} + \cdots \, , 
\end{eqnarray}
where the dots stand for additional terms. 
Terms in eq. (\ref{LagC}), together with the corresponding mass 
terms for ${\bf 15}$ and ${\bf 50}$, produce the $d=9$ operators 
shown in fig. (\ref{fig:MdlCbb}) on the left.

We introduce the shorthand notation
\begin{eqnarray}\label{eq:defsh}
{\boldsymbol{T}}\equiv\left(\mathbf{3},\mathbf{2},\frac{1}{6}\right),
{\boldsymbol{Z}}\equiv\left(\mathbf{6},\mathbf{1},-\frac{2}{3}\right)
\in\mathbf{15} \, , \\ 
{\boldsymbol{R}}\equiv\left(\mathbf{1},\mathbf{1},-2\right),
{\boldsymbol{X}}\equiv\left(\overline{\mathbf{6}},\mathbf{3},-\frac{1}{3}\right),\,
{\boldsymbol{Y}}\equiv\left(\mathbf{6},\mathbf{1},\frac{4}{3}\right)
\in\mathbf{50}\, .
\end{eqnarray}
Differently from model T-I, we will assume that there are a total of
six light states: 2 copies of ${\boldsymbol{T}}$ plus one copy of
${\boldsymbol{R}}$, ${\boldsymbol{X}}$, ${\boldsymbol{Y}}$ and
${\boldsymbol{Z}}$. The Lagrangian in the SM phase then contains the
following terms:
\begin{eqnarray}\label{eq:LagC}
\mathscr{L}_{FFS} & = &
  y_{ijk}^{(4)}L_{i}d_{j}^{c}{\boldsymbol{T}}_{k} 
 +y_{ij}^{(5)}d_{i}^{c}d_{j}^{c}{\boldsymbol{Z}}+y_{ij}^{(6)}e_{i}^{c}e_{j}^{c}{\boldsymbol{R}}
 +y_{ij}^{(7)}Q_{i}Q_{j}{\boldsymbol{X}}+y_{ij}^{(8)}u_{i}^{c}u_{j}^{c}{\boldsymbol{Y}}
 +\textrm{h.c.}\,,\\
\mathscr{L}_{SSS} & = & h_{ij}^{(1)}{\boldsymbol{T}}_{i}{\boldsymbol{T}}_{j}{\boldsymbol{X}}
        +h^{(2)}{\boldsymbol{Y}}{\boldsymbol{Z}}{\boldsymbol{Z}}
        +h^{(3)}{\boldsymbol{R}}^{*}{\boldsymbol{Y}}^{*}{\boldsymbol{Z}}
        +h^{(4)}{\boldsymbol{X}}^{*}{\boldsymbol{X}}^{*}{\boldsymbol{Z}}+\textrm{h.c.}\, .\,
\end{eqnarray}
With these terms the diagrams in the left of fig. (\ref{fig:MdlCbb})
produce two diagrams each in the SM phase of the model. These are
shown in fig. (\ref{fig:MdlCbb}) on the right. We stress again, that
the central couplings in these diagrams, proportional to the couplings
$h^{(1)}-h^{(4)}$, receive contributions from several terms of the
Lagrangian in the $SU(5)$ phase of the model and, as discussed in the
appendix, can therefore take quite different values, despite their
common origin.

Consider first the diagrams on the right-hand side of
fig. (\ref{fig:MdlCbb}) containing two external leptons. We can easily
identify them with the T-II operators classified in
\cite{Bonnet:2012kh} as
\begin{eqnarray}\label{eq:TIIop11}
{\cal O}_{11}^{{\rm T-II}-4} = ({\overline L}{\overline d^c})({\overline L}{\overline d^c})({\overline Q} \overline Q) \,, \\ 
\label{eq:TIIopm}
{\cal O}_{-}^{{\rm T-II}-3} = (e^ce^c)({\overline d^c}{\overline d^c})(u^c u^c) \,.
\end{eqnarray}
Here, again the subscript ``11'' identifies the Babu \& Leung operator
\cite{Babu:2001ex}, whereas ``--'' is the missing $\Delta L=2$
dimension 9 operator in this list. The superscripts indicate the
$\znbb$ decay label in the classification scheme of
\cite{Bonnet:2012kh}. The latter allows us to estimate the $\znbb$
decay rates induced by these two diagrams. By coincidence the two
operators give very similar limits: using again the experimental
limits from \cite{Albert:2014awa,Gando:2012zm,Asakura:2014lma} results
in $M_{\rm eff} \gsim 2.2 g_{\rm eff}^{4/5}$ TeV for each of the two
operators. As before, $M_{\rm eff}$ stands for the mean of the masses
entering the diagram and for the definition of the mean coupling
$g_{\rm eff}$ we have made $h^{(1)}$ and $h^{(3)}$ dimensionless by
introducing $g^{(k)}\equiv h^{(k)}/M_{\rm eff}$. We stress that, in
contrast to model T-I, here all light states appear in at least one
tree-level double beta decay diagram. However, this variant model does
not include a dark matter candidate.

The two diagrams on the right-hand side of fig. (\ref{fig:MdlCbb})
also produce the operators:
\begin{eqnarray}
{\cal O}_{\Delta B=2,\Delta L=0} = ({\overline d^c}{\overline d^c})
                    ({\overline d^c}{\overline d^c})(\overline u^c \overline u^c) \, , \\ 
{\cal O}_{\Delta B=2,\Delta L=0} = (QQ)({\overline d^c}{\overline d^c})(Q Q) \, .
\end{eqnarray}
Both of these operators induce $n-{\bar n}$ oscillations --- see for
example the review \cite{Mohapatra:2009wp}. The current limit on the
life-time of neutron-antineutron oscillations is $\tau_{n\to{\bar n}}
> 0.86 \times 10^8$ s \cite{Phillips:2014fgb} and, as discussed in
that paper, one expects an improvement on this number of about two
orders of magnitude in the future. We can estimate the life-time of
neutron-antineutrino oscillations, $\tau_{n-{\bar n}}$ adapting the
formulas from \cite{Babu:2012vc} for our model. Introducing the
notation ${\overline g} =
(y_{11}^{(7)})^{2/3}(y_{11}^{(5)})^{1/3}$, one gets from the current 
experimental limit a constraint 
\begin{equation}\label{eq:limh4}
h^{(4)}\lsim0.6\times10^{-6}\left(\frac{0.1}{{\overline{g}}}\right)^{3}\left(\frac{M_{\rm eff}}{3\hskip1mm{\rm TeV}}\right)^{6}\hskip2mm{\rm GeV}
\end{equation}
for the bottom right diagram in fig. (\ref{fig:MdlCbb}) (in this case,
$M_{\rm eff}=(m_X^4m_Z^2)^{1/6}$). The same constraint applies to
$h^{(2)}$ from the top right diagram of fig. (\ref{fig:MdlCbb}). This
fine-tuning could only be avoided if $\boldsymbol Z$ takes a mass of
the order of the GUT scale, as in the model of \cite{Babu:2012vc},
which would, however, destroy gauge coupling unification in model T-II
discussed next.

\begin{figure}[t]
\includegraphics[scale=0.70]{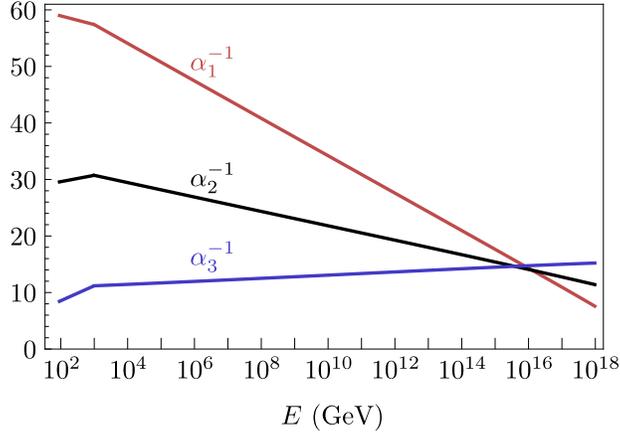}
\caption{Gauge coupling unification in model T-II. In this example
  calculation the new states are assumed to have masses around $M \sim
  {\cal O}(1)$ TeV. As in fig. (\ref{fig:gcuA}), the running includes
  2-loop $\beta$ coefficients, but does not consider any GUT scale
  thresholds.}
\label{fig:gcuC}
\end{figure}

For model T-II we find the following $\beta$ coefficients:
\begin{eqnarray}
\label{biMC}
b_i=
\begin{pmatrix}
\frac{241}{30} \\  \frac{11}{6} \\ -\frac{13}{6} 
\end{pmatrix}
& , \qquad &
b_{ij} =
\begin{pmatrix}
\frac{1691}{30} & \frac{129}{10} & \frac{1988}{15} \\
\frac{43}{10} & \frac{785}{6} & 188 \\ 
\frac{497}{30} & \frac{141}{2} & \frac{541}{3} 
\end{pmatrix}\, .
\end{eqnarray}
Fig. (\ref{fig:gcuC}) shows the corresponding running of the inverse
gauge couplings for masses of the new particles $M \sim {\cal O}(1)$
TeV. Despite the very different particle content of model T-II,
compared to model T-I, GCU works very nicely. The estimated GUT scale
in this model is found to be roughly $m_G \simeq 5 \times 10^{15}$
GeV, lower than in model T-I. Also $\alpha_G$ is predicted to be much
larger than in model T-I. This can be easily understood from the
$\beta$ coefficients, which are larger due to the presence of colour
sextets in the running.

The best-fit point for the proton decay half-life is found to be
$\tau_{p\to\pi^{0}e^{+}}\simeq 5 \times 10^{33}$ years, slightly below
the current experimental limit \cite{Abe:2013lua}. However, it is
well-known that predictions of proton decay half-lives have a large
error bar. We have repeated the exercise of estimating the uncertainty
in the GUT scale due to the neglected GUT-scale thresholds, following
the $\chi^2$ procedure of \cite{Arbelaez:2013nga}.  We will not show
the plots corresponding to model T-II here, since the results are
similar to the example models discussed in \cite{Arbelaez:2013nga}.
From this exercise we estimate that typical uncertainties in the
predicted proton-decay half-life should be around ($2-2.5$) orders of
magnitude. The planned Hyper-Kamiokande experiment might cover a large
part of this range \cite{Abe:2011ts}.  Note that, since model T-II has
the same light leptoquark state ${\boldsymbol{T}}$ which appears in
model T-I, the constraint on scalar mediated proton decay, see
eq. (\ref{eq:limh1}), applies also in model T-II.

\begin{figure}[t]
\includegraphics[scale=0.95]{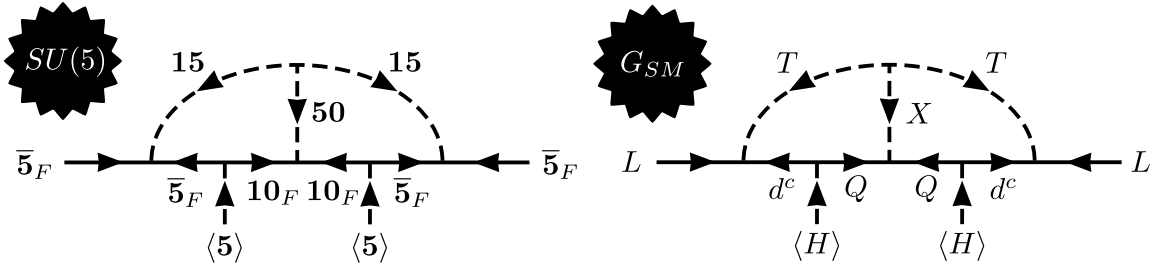}
\caption{2-loop neutrino mass diagrams in model T-II.
The structure is as in BZ}
\label{fig:mnuC}
\end{figure}

We now briefly discuss neutrino masses. Since the GUT scale in model T-II
is expected to be lower than in model T-I, the type-II seesaw
contribution could be as large as $(m_{\nu})_{ij} \simeq 6
y_{ij}^{(4)}$ meV, assuming $\mu_{\Delta}=m_{\Delta} \simeq
m_G$. This is, in principle, large enough to explain the solar
neutrino scale. However, $\mu_{\Delta} \ll m_G$ and/or
$y_{ij}^{(4)} < 1$ would render this contribution negligible.

More important is the radiative contribution from TeV scale particles.
The operators in eqs. (\ref{eq:TIIop11}) and (\ref{eq:TIIopm}) will
produce 2-loop (${\cal O}_{11}^{{\rm T-II}-4}$) and 4-loop (${\cal
  O}_{-}^{{\rm T-II}-3}$) neutrino masses. The contribution to the
neutrino mass matrix from 4-loop diagrams is expected to be at most
$(m_{\nu})_{\tau\tau} \sim {\cal O}(10^{-10})$ eV \cite{Helo:2015fba},
so completely negligible numerically. The 2-loop diagram shown in
fig. (\ref{fig:mnuC}), on the other hand, will produce neutrino masses
of the same order of eq. (\ref{eq:estmnu}) \cite{Sierra:2014rxa}.  We
do not repeat the discussion of fits to neutrino data here, since it
is very similar to that of the previous subsection. Also in model T-II
LFV rates are expected to be near experimental limits.

We turn now to the discussion of LHC searches. For the constraints on
the leptoquark, ${\boldsymbol{T}}$, see the discussion for model
T-I. The same constraints apply, of course, also in model T-II.  Cross
sections for colour sextets are expected to be particularly large at
the LHC \cite{Han:2010rf}, since they can be produced as s-channel
resonances. Both, ATLAS \cite{Aad:2014aqa} and CMS
\cite{Khachatryan:2015sja} have searched for the appearance of new
resonances in dijet spectra. Especially, \cite{Khachatryan:2015sja}
presents upper limits on $\sigma(pp\to jj)$ as function of the mass of
a hypothetical resonance coupled to pairs of quarks.  We can combine
these limits with cross section calculations for resonances coupling
to $uu$ or $dd$ pairs, see for example \cite{Helo:2013ika}. We
estimate that, for $m_Y$ ($m_Z$) equal to 3 TeV this results in upper
limits on the couplings $y_{11}^{(8)}$ ($y_{11}^{(5)}$) of roughly
$y_{11}^{(8)}\lsim 0.19$ ($y_{11}^{(5)} \lsim 0.056$). Similar bounds
can be derived for $\boldsymbol{X}$ (which nevertheless is an
$SU(2)_L$ triplet, unlike $\boldsymbol{Y}$ and $\boldsymbol{Z}$).
Run-2 of the LHC will improve vastly on these numbers.

Much smaller couplings (and larger masses) can be probed at the LHC,
if there is a certain hierarchy in masses among the states ${\bf X}$,
${\boldsymbol{Y}}$, ${\boldsymbol{Z}}$, ${\boldsymbol{R}}$ and
${\boldsymbol{T}}$.  We will discuss one example. If $m_X > 2 m_T$,
then the diquark ${\boldsymbol{X}}$ will have a non-negligible decay
rate to a pair of leptoquarks --- see the top-right diagram in
fig. (\ref{fig:MdlCbb}).  The branching ratio Br(${\boldsymbol{X}} \to
{\boldsymbol{T}}{\boldsymbol{T}}$) depends on the relative size of the
couplings $h^{(1)}/m_T$ and $y_{ij}^{(7)}$.  If the branching ratio is
non-zero, since each of the ${\boldsymbol{T}}$'s will decay into
$l^{+}j$, the total signal is $l^{+}l^{+}jj$, which enjoys a
background orders of magnitude lower than the dijet spectrum. Since
there is little or no background in this search for large values of
$m_X$, it might be possible to establish discovery with as few a
(3--5) events.

Given that we have no theory for the couplings, $h^{(1)}/m_T$ could be 
much smaller than $y_{ij}^{(7)}$. However, double beta decay depends 
on the same couplings. Thus, a measured finite half-life in a future 
$\znbb$ decay experiment would define a lower limit on 
$h^{(1)}$ as a function of the mass $M_{\rm eff}$ in a parameter range 
that should be completely coverable in the run-II of the LHC, 
considering the large diquark cross sections.

Sensitivity for the doubly charged scalars at the LHC is much weaker,
because of their colour singlet nature. Results of searches for
same-sign dilepton pairs in electron and muon final states have been
published by ATLAS \cite{ATLAS:2014kca}. No deviation from SM
expectations have been found and lower limits on the mass of
${\boldsymbol{R}}$ of roughly $m_R \gsim 400$ GeV are derived for
${\boldsymbol{R}}$ decaying to either $e^{\pm}e^{\pm}$,
$e^{\pm}\mu^{\pm}$ or $\mu^{\pm}\mu^{\pm}$ with 100\% branching
ratio. In summary, model T-II has a particularly rich phenomenology at
the LHC.

\section{Conclusions}

If $B-L$ violation exists at the TeV scale, one expects also sizeable
contributions to the short-range part of the $\znbb$ decay amplitude
\cite{Pas:2000vn,Bonnet:2012kh}, with interesting consequences for
the viability of leptogenesis \cite{Deppisch:2013jxa,Deppisch:2015yqa}.
In this paper we have studied top-down scenarios inspired by $SU(5)$
unification. These models violate $B-L$, thus generate neutrino masses
and neutrino oscillation data can be easily fitted. As in all
TeV-scale models of loop neutrino masses, one expects that charged
lepton flavour violation is large and possibly within reach of near
future experiments. The presence of new states at the
TeV scale changes the running of the three gauge couplings 
in such a way that, unlike in the standard model, gauge coupling 
unification can be achieved without introducing supersymmetry.

We have constructed two exemplary models, one for each of the two
tree-level topologies of $\znbb$ decay. Model T-I has a dark matter 
candidate and few low-scale (TeV) particles. On the other hand, there is 
model T-II which introduces
several new scalars and thus has a richer phenomenology.  The two
models differ in their predicted gauge mediated proton decay
half-lives, due to their different GUT scales. Model T-II might be in
reach of the planned Hyper-Kamiokande experiment \cite{Abe:2011ts}.

The new states predicted by our models can be searched for at the
LHC. We have discussed several different possible LHC searches, the
most interesting of which are certainly the lepton number violating final
states $l^{\pm}l^{\pm}+2 j$ and $l^{\pm}l^{\pm}+4 j$, where
$l=e,\mu,\tau$.  LHC can probe the part of parameter space of our
models where the new states give the dominant contribution to the
$\znbb$ decay amplitude. In this sense, our models are falsifiable 
experimentally. We are looking forward to the results of run-II 
at the LHC.

\bigskip

\centerline{\bf Acknowledgements}

\medskip

The authors thank S.G. Kovalenko for discussions.
Work supported by MINECO grants FPA2014-58183-P, Multidark CSD2009-00064
and the PROMETEOII/2014/084 grant from Generalitat Valenciana.

\section{Appendices}

\subsection*{\label{sec:AppendixA}A -- Lagrangians of the models}

This appendix contains the mass and interaction terms of the two models
discussed in the main text.

\subsubsection*{Model T-I}

Consider the usual three copies of the left-handed fermion representations
$\overline{\mathbf{5}}_{F}$ and $\mathbf{10}_{F}$ of $SU(5)$ in
addition to one $\mathbf{24}_{F}$. With an extended scalar sector
consisting of one copy of the complex representations $\mathbf{5}$,
$\mathbf{15}$, $\mathbf{45}$, $\mathbf{70}$ and a $\mathbf{24}$
real scalar field, the allowed mass and interaction terms are the
following:\thinmuskip=0.5mu
\medmuskip=1mu 
\thickmuskip=1.5mu
\begin{align}
\widehat{\mathscr{L}}_{\textrm{int}} & =\widehat{\mathscr{L}}_{FF}+\widehat{\mathscr{L}}_{FFS}+\widehat{\mathscr{L}}_{SS}+\widehat{\mathscr{L}}_{SSS}+\widehat{\mathscr{L}}_{SSSS}\,,\\
\widehat{\mathscr{L}}_{FF} & =\textrm{\ensuremath{m_{\mathbf{24}}\mathbf{24}_{F}\mathbf{24}_{F}}}+\textrm{h.c.}\,,\\
\widehat{\mathscr{L}}_{FFS} & =\widehat{y}_{ij}^{(1)}\overline{\mathbf{5}}_{F,i}\mathbf{10}_{F,j}\mathbf{5}^{*}+\widehat{y}_{ij}^{(2)}\mathbf{10}_{F,i}\mathbf{10}_{F,j}\mathbf{5}+\widehat{y}_{ij}^{(3)}\overline{\mathbf{5}}_{F,i}\mathbf{10}_{F,j}\mathbf{45}^{*}+\widehat{y}_{ij}^{(4)}\mathbf{10}_{F,i}\mathbf{10}_{F,j}\mathbf{45}\nonumber \\
 & +\widehat{y}_{ij}^{(5)}\overline{\mathbf{5}}_{F,i}\overline{\mathbf{5}}_{F,j}\mathbf{15}+\widehat{y}_{i}^{(6)}\overline{\mathbf{5}}_{F,i}\mathbf{24}_{F}\mathbf{5}+\widehat{y}_{i}^{(7)}\mathbf{10}_{F,i}\mathbf{24}_{F}\mathbf{15}^{*}+\widehat{y}_{i}^{(8)}\overline{\mathbf{5}}_{F,i}\mathbf{24}_{F}\mathbf{45}\nonumber \\
 & +\widehat{y}_{i}^{(9)}\overline{\mathbf{5}}_{F,i}\boldsymbol{24}_{F}\mathbf{70}+\widehat{y}^{(10)}\boldsymbol{24}_{F}\boldsymbol{24}_{F}\mathbf{24}+\textrm{h.c.}\,,\\
\widehat{\mathscr{L}}_{SS} & =m_{\mathbf{5}}^{2}\mathbf{5}\cdot\mathbf{5}^{*}+m_{\mathbf{15}}^{2}\mathbf{15}\cdot\mathbf{15}^{*}+m_{\mathbf{45}}^{2}\mathbf{45}\cdot\mathbf{45}^{*}+m_{\mathbf{50}}^{2}\mathbf{70}\cdot\mathbf{70}^{*}+\frac{1}{2}m_{\mathbf{24}}^{2}\mathbf{24}\cdot\mathbf{24}\,,\\
\widehat{\mathscr{L}}_{SSS} & =\left(\widehat{h}^{(1)}\mathbf{5}\cdot\mathbf{5}\cdot\mathbf{15}^{*}+\widehat{h}^{(2)}\mathbf{5}\cdot\mathbf{15}^{*}\cdot\mathbf{70}+\widehat{h}^{(3)}\mathbf{5}\cdot\mathbf{24}\cdot\mathbf{45}^{*}+\widehat{h}^{(4)}\mathbf{5}\cdot\mathbf{24}\cdot\mathbf{70}^{*}+\widehat{h}^{(5)}\mathbf{15}\cdot\mathbf{45}^{*}\cdot\mathbf{45}^{*}\right.\nonumber \\
 & \left.+\widehat{h}^{(6)}\mathbf{15}\cdot\mathbf{45}^{*}\cdot\mathbf{70}^{*}+\widehat{h}^{(7)}\mathbf{15}\cdot\mathbf{70}^{*}\cdot\mathbf{70}^{*}+\widehat{h}^{(8)}\mathbf{24}\cdot\mathbf{45}\cdot\mathbf{70}^{*}+\textrm{h.c.}\right)+\widehat{h}^{(9)}\mathbf{5}\cdot\mathbf{5}^{*}\cdot\mathbf{24}\nonumber \\
 & +\widehat{h}^{(10)}\mathbf{15}\cdot\mathbf{15}^{*}\cdot\mathbf{24}+\widehat{h}^{(11;a)}\left[\mathbf{24}\cdot\mathbf{45}\cdot\mathbf{45}^{*}\right]_{a=1,2}+\widehat{h}^{(12;a)}\left[\mathbf{70}\cdot\mathbf{70}^{*}\cdot\mathbf{24}\right]_{a=1,2}+\widehat{h}^{(13)}\mathbf{24}\cdot\mathbf{24}\cdot\mathbf{24}\,,\label{eq:5}\\
\widehat{\mathscr{L}}_{SSSS} & =\widehat{\lambda}^{(1)}\mathbf{5}\cdot\mathbf{5}\cdot\mathbf{5}^{*}\cdot\mathbf{70^{*}}+\widehat{\lambda}^{(2)}\mathbf{5}\cdot\mathbf{5}\cdot\mathbf{45}^{*}\cdot\mathbf{70^{*}}+\widehat{\lambda}^{(3)}\mathbf{5}\cdot\mathbf{5}^{*}\cdot\mathbf{45}\cdot\mathbf{70^{*}}+\widehat{\lambda}^{(4;a)}\left[\mathbf{5}\cdot\mathbf{45}\cdot\mathbf{45}^{*}\cdot\mathbf{70^{*}}\right]_{a=1,2}\nonumber \\
 & +\widehat{\lambda}^{(5;a)}\left[\mathbf{5}^{*}\cdot\mathbf{45}\cdot\mathbf{45}\cdot\mathbf{70^{*}}\right]_{a=1,\cdots,4}+\widehat{\lambda}^{(6;a)}\left[\mathbf{45}\cdot\mathbf{45}\cdot\mathbf{45}^{*}\cdot\mathbf{70^{*}}\right]_{a=1,\cdots,5}+\left(\textrm{other terms}\right)\,,
\end{align}
\thinmuskip=1mu
\medmuskip=4mu plus 2mu minus 4mu
\thickmuskip=5mu plus 5mu where the Yukawa couplings $\widehat{y}_{ij}^{(2)}$ and $\widehat{y}_{ij}^{(5)}$
are symmetric under an exchange of the $\left(i,j\right)$ indices.
The notation $\left[\cdots\right]_{a=1,2,\cdots}$ is used above to
indicate the existence of multiple gauge invariant contractions of
a given product of fields. Quartic scalar couplings are numerous,
therefore we do not write them all down here. The ones shown are important
because, in their absence, it is possible to have a viable dark model
candidate (which is the $\boldsymbol{K}$ scalar field defined in
the following; see the main text). We also note in passing that, as
usual in $SU(5)$, the scalar $\mathbf{5}$ and $\mathbf{45}$ are
necessary to generate the Yukawa couplings $\widehat{y}_{ij}^{(1)},\cdots,\widehat{y}_{ij}^{(4)}$
which in turn will allow the SM fermions to have realistic couplings
to the Higgs field (which is a combination of the $SU(2)$ doublets
found inside the $\mathbf{5}$ and $\mathbf{45}$).

The $SU(5)$ gauge symmetry breaks down into $SU(3)\times SU(2)\times U(1)$
once the scalar $\mathbf{24}$ acquires a VEV in the singlet component
(under the SM group). Using the same notation as in the main text,
the fields which remain light --- besides the SM fermions --- are
\begin{gather}
H\equiv\left(\mathbf{1},\mathbf{2},\frac{1}{2}\right)\in\mathbf{5}\textrm{ and }\mathbf{45}\,,\quad\boldsymbol{O}\equiv\left(\mathbf{8},\mathbf{1},0\right)\in\mathbf{24}_{F}\,,\\
\boldsymbol{T}\equiv\left(\mathbf{3},\mathbf{2},\frac{1}{6}\right)\in\mathbf{15}\,,\quad\boldsymbol{K}\equiv\left(\mathbf{1},\mathbf{4},\frac{1}{2}\right)\in\mathbf{70}\,.
\end{gather}
The most general form of the interaction Lagrangian (including mass
terms) which one can form with these fields is as follows:\thinmuskip=0.5mu
\medmuskip=1mu 
\thickmuskip=1.5mu
\begin{align}
\mathscr{L}_{\textrm{int}} & =\mathscr{L}_{FF}+\mathscr{L}_{FFS}+\mathscr{L}_{SS}+\mathscr{L}_{SSS}+\mathscr{L}_{SSSS}\,,\\
\mathscr{L}_{FF} & =m_{O}\boldsymbol{O}\boldsymbol{O}\,,\\
\mathscr{L}_{FFS} & =y_{ij}^{(1)}Q_{i}u_{j}^{c}H+y_{ij}^{(2)}Q_{i}d_{j}^{c}H^{*}+y_{ij}^{(3)}L_{i}e_{j}^{c}H^{*}+y_{ij}^{(4)}L_{i}d_{j}^{c}\boldsymbol{T}+y_{i}^{(5)}Q_{i}\boldsymbol{O}\boldsymbol{T}^{*}+\textrm{h.c.}\,,\\
\mathscr{L}_{SS} & =m_{H}^{2}HH^{*}+m_{T}^{2}\boldsymbol{T}\boldsymbol{T}^{*}+m_{K}^{2}\boldsymbol{K}\boldsymbol{K}^{*}\,,\\
\mathscr{L}_{SSS} & =0\,,\\
\mathscr{L}_{SSSS} & =\left(\lambda^{(1)}HHH^{*}\boldsymbol{K}^{*}+\lambda^{(2)}HH\boldsymbol{K}^{*}\boldsymbol{K}^{*}+\lambda^{(3)}H\boldsymbol{T}\boldsymbol{T}^{*}\boldsymbol{K}^{*}+\lambda^{(4)}H\boldsymbol{K}\boldsymbol{K}^{*}\boldsymbol{K}^{*}+\textrm{h.c.}\right)\nonumber \\
 & +\lambda^{(5)}HHH^{*}H^{*}+\lambda^{(6;a)}\left[HH^{*}\boldsymbol{T}\boldsymbol{T}^{*}\right]_{a=1,2}+\lambda^{(7;a)}\left[HH^{*}\boldsymbol{K}\boldsymbol{K}^{*}\right]_{a=1,2}\nonumber \\
 & +\lambda^{(8;a)}\left[\boldsymbol{T}\boldsymbol{T}\boldsymbol{T}^{*}\boldsymbol{T}^{*}\right]_{a=1,2}+\lambda^{(9;a)}\left[\boldsymbol{T}\boldsymbol{T}^{*}\boldsymbol{K}\boldsymbol{K}^{*}\right]_{a=1,2}+\lambda^{(10;a)}\left[\boldsymbol{K}\boldsymbol{K}\boldsymbol{K}^{*}\boldsymbol{K}^{*}\right]_{a=1,2}\,.
\end{align}
\thinmuskip=1mu
\medmuskip=4mu plus 2mu minus 4mu
\thickmuskip=5mu plus 5mu

\subsubsection*{Model T-II}

This model contains, as usual, three copies of the left-handed fermion
representations $\overline{\mathbf{5}}_{F}$ and $\mathbf{10}_{F}$
of $SU(5)$. In addition, there are complex scalars $\mathbf{5}$,
$\mathbf{45}$, $\mathbf{50}$, two copies of the complex scalar $\mathbf{15}$,
and one copy of the real scalars $\mathbf{24}$ and $\mathbf{75}$.
With this field content, we can write the following terms:\thinmuskip=0.5mu
\medmuskip=1mu 
\thickmuskip=1.5mu
\begin{align}
\mathscr{\widehat{L}}_{\textrm{int}} & =\mathscr{\widehat{L}}_{FFS}+\mathscr{\widehat{L}}_{SS}+\mathscr{\widehat{L}}_{SSS}+\mathscr{\widehat{L}}_{SSSS}\,,\\
\mathscr{\widehat{L}}_{FFS} & =\widehat{y}_{ij}^{(1)}\overline{\mathbf{5}}_{F,i}\mathbf{10}_{F,j}\mathbf{5}^{*}+\widehat{y}_{ij}^{(2)}\mathbf{10}_{F,i}\mathbf{10}_{F,j}\mathbf{5}+\widehat{y}_{ij}^{(3)}\overline{\mathbf{5}}_{F,i}\mathbf{10}_{F,j}\mathbf{45}^{*}+\widehat{y}_{ij}^{(4)}\mathbf{10}_{F,i}\mathbf{10}_{F,j}\mathbf{45}\nonumber \\
 & +\widehat{y}_{ijk}^{(5)}\overline{\mathbf{5}}_{F,i}\overline{\mathbf{5}}_{F,j}\mathbf{15}_{k}+\widehat{y}_{ij}^{(6)}\mathbf{10}_{F,i}\mathbf{10}_{F,j}\mathbf{50}+\textrm{h.c.}\,,\\
\mathscr{\widehat{L}}_{SS} & =m_{\mathbf{5}}^{2}\mathbf{5}\cdot\mathbf{5}^{*}+\left(m_{\mathbf{15}}^{2}\right)_{ij}\mathbf{15}_{i}\cdot\mathbf{15}_{j}^{*}+m_{\mathbf{45}}^{2}\mathbf{45}\cdot\mathbf{45}^{*}+m_{\mathbf{50}}^{2}\mathbf{50}\cdot\mathbf{50}^{*}+\frac{1}{2}m_{\mathbf{24}}^{2}\mathbf{24}\cdot\mathbf{24}+\frac{1}{2}m_{\mathbf{75}}^{2}\mathbf{75}\cdot\mathbf{75}\,,\\
\mathscr{\widehat{L}}_{SSS} & =\left(\widehat{h}_{i}^{(1)}\mathbf{5}\cdot\mathbf{5}\cdot\mathbf{15}_{i}^{*}+\widehat{h}_{ij}^{(2)}\mathbf{15}_{i}\cdot\mathbf{15}_{j}\cdot\mathbf{50}+\widehat{h}_{i}^{(3)}\mathbf{15}_{i}\cdot\mathbf{50}^{*}\cdot\mathbf{50}^{*}+\widehat{h}^{(4)}\mathbf{5}\cdot\mathbf{24}\cdot\mathbf{45}^{*}+\widehat{h}_{i}^{(5)}\mathbf{15}_{i}\cdot\mathbf{45}^{*}\cdot\mathbf{45}^{*}\right.\nonumber \\
 & +\widehat{h}^{(6;a)}\left[\mathbf{24}\cdot\mathbf{45}\cdot\mathbf{45}^{*}\right]_{a=1,2}+\widehat{h}^{(7)}\mathbf{24}\cdot\mathbf{45}\cdot\mathbf{50}^{*}+\widehat{h}^{(8)}\mathbf{5}\cdot\mathbf{45}^{*}\cdot\mathbf{75}+\widehat{h}^{(9)}\mathbf{5}\cdot\mathbf{50}^{*}\cdot\mathbf{75}\nonumber \\
 & \left.+\widehat{h}^{(10)}\mathbf{45}\cdot\mathbf{50}^{*}\cdot\mathbf{75}+\textrm{h.c.}\right)+\widehat{h}^{(11)}\mathbf{5}\cdot\mathbf{5}^{*}\cdot\mathbf{24}+\widehat{h}_{ij}^{(12)}\mathbf{15}_{i}\cdot\mathbf{15}_{j}^{*}\cdot\mathbf{24}+\widehat{h}^{(13)}\mathbf{24}\cdot\mathbf{24}\cdot\mathbf{24}\nonumber \\
 & +\widehat{h}^{(14)}\mathbf{50}\cdot\mathbf{50}^{*}\cdot\mathbf{24}+\widehat{h}^{(15;a)}\left[\mathbf{45}\cdot\mathbf{45}^{*}\cdot\mathbf{75}\right]_{a=1,2}+\widehat{h}^{(16)}\mathbf{50}\cdot\mathbf{50}^{*}\cdot\mathbf{75}+\widehat{h}^{(17)}\mathbf{24}\cdot\mathbf{24}\cdot\mathbf{75}\nonumber \\
 & +\widehat{h}^{(18)}\mathbf{24}\cdot\mathbf{75}\cdot\mathbf{75}+\widehat{h}^{(19)}\mathbf{75}\cdot\mathbf{75}\cdot\mathbf{75}\,,\label{eq:18}\\
\mathscr{\widehat{L}}_{SSSS} & =\widehat{\lambda}_{ij}^{(1;a)}\left[\mathbf{15}_{i}\cdot\mathbf{15}_{j}\cdot\mathbf{50}\cdot\mathbf{24}\right]_{a=1,2}+\widehat{\lambda}_{ij}^{(2)}\mathbf{15}_{i}\cdot\mathbf{15}_{j}\cdot\mathbf{50}\cdot\mathbf{75}+\widehat{\lambda}_{i}^{(3)}\mathbf{15}_{i}\cdot\mathbf{50}^{*}\cdot\mathbf{50}^{*}\cdot\mathbf{24}\nonumber \\
 & +\widehat{\lambda}_{i}^{(4;a)}\left[\mathbf{15}_{i}\cdot\mathbf{50}^{*}\cdot\mathbf{50}^{*}\cdot\mathbf{75}\right]_{a=1,2}+\widehat{\lambda}_{i}^{(5;a)}\left[\mathbf{15}_{i}\cdot\mathbf{15}_{j}^{*}\cdot\mathbf{24}\cdot\mathbf{24}\right]_{a=1,2,3}+\left(\textrm{other terms}\right)\,.\label{eq:19}
\end{align}
\thinmuskip=1mu
\medmuskip=4mu plus 2mu minus 4mu
\thickmuskip=5mu plus 5muUnder an exchange of the $\left(i,j\right)$ flavour indices $\widehat{y}_{ij}^{(2)}$,
$\widehat{y}_{ij}^{(5)}$, $\widehat{y}_{ij}^{(6)}$, $\widehat{h}_{ij}^{(2)}$,
$\widehat{\lambda}_{ij}^{(1;1)}$ and $\widehat{\lambda}_{ij}^{(2)}$
symmetric, while $\widehat{y}_{ij}^{(4)}$ and $\widehat{\lambda}_{ij}^{(1;2)}$
are anti-symmetric. As explained in appendix \hyperref[sec:AppendixB]{B},
the quartic couplings shown above are necessary in order to suppress
neutron--antineutron oscillations in model T-II.

With the fields
\begin{gather}
H\equiv\left(\mathbf{1},\mathbf{2},\frac{1}{2}\right)\in\mathbf{5}\textrm{ and }\mathbf{45}\,,\quad\boldsymbol{T}\equiv\left(\mathbf{3},\mathbf{2},\frac{1}{6}\right)\,,\quad\boldsymbol{Z}\equiv\left(\mathbf{6},\mathbf{1},-\frac{2}{3}\right)\in\mathbf{15}\,,\\
\boldsymbol{R}\equiv\left(\mathbf{1},\mathbf{1},-2\right)\,,\quad\boldsymbol{X}\equiv\left(\overline{\mathbf{6}},\mathbf{3},-\frac{1}{3}\right)\,,\quad\boldsymbol{Y}\equiv\left(\mathbf{6},\mathbf{1},\frac{4}{3}\right)\in\mathbf{50}\,,
\end{gather}
which are assumed to remain light fields after of $SU(5)$ symmetry
breaking (see appendix \hyperref[sec:AppendixB]{B}), one can build
the following terms (note that there are two light $\boldsymbol{T}$'s
but only one $\boldsymbol{Z}$):
\begin{align}
\mathscr{L}_{\textrm{int}} & =\mathscr{L}_{FFS}+\mathscr{L}_{SS}+\mathscr{L}_{SSS}+\mathscr{L}_{SSSS}\,,\\
\mathscr{L}_{FFS} & =y_{ij}^{(1)}Q_{i}u_{j}^{c}H+y_{ij}^{(2)}Q_{i}d_{j}^{c}H^{*}+y_{ij}^{(3)}L_{i}e_{j}^{c}H^{*}+y_{ijk}^{(4)}L_{i}d_{j}^{c}\boldsymbol{T}_{k}\nonumber \\
 & +y_{ij}^{(5)}d_{i}^{c}d_{j}^{c}\boldsymbol{Z}+y_{ij}^{(6)}e_{i}^{c}e_{j}^{c}\boldsymbol{R}+y_{ij}^{(7)}Q_{i}Q_{j}\boldsymbol{X}+y_{ij}^{(8)}u_{i}^{c}u_{j}^{c}\boldsymbol{Y}+\textrm{h.c.}\,,\\
\mathscr{L}_{SS} & =m_{H}^{2}HH^{*}+\left(m_{T}^{2}\right)_{ij}\boldsymbol{T}_{i}\boldsymbol{T}_{j}^{*}+m_{Z}^{2}\boldsymbol{Z}\boldsymbol{Z}^{*}+m_{R}^{2}\boldsymbol{R}\boldsymbol{R}^{*}+m_{X}^{2}\boldsymbol{X}\boldsymbol{X}^{*}+m_{Y}^{2}\boldsymbol{Y}\boldsymbol{Y}^{*}\,,\\
\mathscr{L}_{SSS} & =h_{ij}^{(1)}\boldsymbol{T}_{i}\boldsymbol{T}_{j}\boldsymbol{X}+h^{(2)}\boldsymbol{Y}\boldsymbol{Z}\boldsymbol{Z}+h^{(3)}\boldsymbol{R}^{*}\boldsymbol{Y}^{*}\boldsymbol{Z}+h^{(4)}\boldsymbol{X}^{*}\boldsymbol{X}^{*}\boldsymbol{Z}+\textrm{h.c.}\,,\\
\mathscr{L}_{SSSS} & =\left(\lambda^{(1)}HH\boldsymbol{X}\boldsymbol{Z}+\lambda^{(2)}HH\boldsymbol{X}^{*}\boldsymbol{Y}^{*}+\lambda_{ijk}^{(3;a)}\left[H\boldsymbol{T}_{i}^{*}\boldsymbol{T}_{j}^{*}\boldsymbol{T}_{k}^{*}\right]_{a=1,2}+\lambda_{ij}^{(4)}\boldsymbol{T}_{i}\boldsymbol{T}_{j}\boldsymbol{X}^{*}\boldsymbol{Z}\right.\nonumber \\
 & \left.+\lambda^{(5)}\boldsymbol{R}^{*}\boldsymbol{Z}\boldsymbol{Z}\boldsymbol{Z}+\lambda^{(6)}\boldsymbol{R}\boldsymbol{Y}\boldsymbol{Y}\boldsymbol{Z}+\lambda^{(7;a)}\left[\boldsymbol{X}\boldsymbol{X}\boldsymbol{Y}\boldsymbol{Z}\right]_{a=1,2}+\lambda^{(8)}\boldsymbol{R}\boldsymbol{X}^{*}\boldsymbol{X}^{*}\boldsymbol{Y}+\textrm{h.c.}\right)\nonumber \\
 & +\lambda^{(9)}HHH^{*}H^{*}+\lambda_{ij}^{(10)}HH^{*}\boldsymbol{T}_{i}\boldsymbol{T}_{j}^{*}+\lambda^{(11)}HH^{*}\boldsymbol{Z}\boldsymbol{Z}^{*}+\lambda^{(12)}HH^{*}\boldsymbol{R}\boldsymbol{R}^{*}\nonumber \\
 & +\lambda^{(13;a)}\left[HH^{*}\boldsymbol{X}\boldsymbol{X}^{*}\right]_{a=1,2}+\lambda^{(14)}HH^{*}\boldsymbol{Y}\boldsymbol{Y}^{*}+\lambda_{ijkl}^{(15;a)}\left[\boldsymbol{T}_{i}\boldsymbol{T}_{j}\boldsymbol{T}_{k}^{*}\boldsymbol{T}_{l}^{*}\right]_{a=1,\cdots,4}\nonumber \\
 & +\lambda_{ij}^{(16;a)}\left[\boldsymbol{T}_{i}\boldsymbol{T}_{j}^{*}\boldsymbol{Z}\boldsymbol{Z}^{*}\right]_{a=1,2}+\lambda_{ij}^{(17)}\boldsymbol{R}\boldsymbol{R}^{*}\boldsymbol{T}_{i}\boldsymbol{T}_{j}^{*}+\lambda_{ij}^{(18;a)}\left[\boldsymbol{T}_{i}\boldsymbol{T}_{j}^{*}\boldsymbol{X}\boldsymbol{X}^{*}\right]_{a=1,\cdots,4}\nonumber \\
 & +\lambda_{ij}^{(19;a)}\left[\boldsymbol{T}_{i}\boldsymbol{T}_{j}^{*}\boldsymbol{Y}\boldsymbol{Y}^{*}\right]_{a=1,2}+\lambda^{(20;a)}\left[\boldsymbol{Z}\boldsymbol{Z}\boldsymbol{Z}^{*}\boldsymbol{Z}^{*}\right]_{a=1,2}+\lambda^{(21)}\boldsymbol{R}\boldsymbol{R}^{*}\boldsymbol{Z}\boldsymbol{Z}^{*}\nonumber \\
 & +\lambda^{(22;a)}\left[\boldsymbol{X}\boldsymbol{X}^{*}\boldsymbol{Z}\boldsymbol{Z}^{*}\right]_{a=1,2,3}+\lambda^{(23;a)}\left[\boldsymbol{Y}\boldsymbol{Y}^{*}\boldsymbol{Z}\boldsymbol{Z}^{*}\right]_{a=1,2,3}+\lambda^{(24)}\boldsymbol{R}\boldsymbol{R}\boldsymbol{R}^{*}\boldsymbol{R}^{*}\nonumber \\
 & +\lambda^{(25)}\boldsymbol{R}\boldsymbol{R}^{*}\boldsymbol{X}\boldsymbol{X}^{*}+\lambda^{(26)}\boldsymbol{R}\boldsymbol{R}^{*}\boldsymbol{Y}\boldsymbol{Y}^{*}+\lambda^{(27;a)}\left[\boldsymbol{X}\boldsymbol{X}\boldsymbol{X}^{*}\boldsymbol{X}^{*}\right]_{a=1,\cdots,5}\nonumber \\
 & +\lambda^{(28;a)}\left[\boldsymbol{X}\boldsymbol{X}^{*}\boldsymbol{Y}\boldsymbol{Y}^{*}\right]_{a=1,2,3}+\lambda^{(29;a)}\left[\boldsymbol{Y}\boldsymbol{Y}\boldsymbol{Y}^{*}\boldsymbol{Y}^{*}\right]_{a=1,2}\,.
\end{align}
There are the following symmetries in the parameters:
\begin{itemize}
\item $\widehat{y}_{ij}^{(5)}$, $\widehat{y}_{ij}^{(6)}$, $\widehat{y}_{ij}^{(7)}$,
$\widehat{y}_{ij}^{(8)}$, $\widehat{h}_{ij}^{(1)}$ and $\widehat{\lambda}_{ij}^{(4)}$
are symmetric under an exchange $i\leftrightarrow j$.
\item $\widehat{\lambda}_{ijk}^{(3;1)}$ and $\widehat{\lambda}_{ijk}^{(3;2)}$
have a mixed symmetry under permutations of the three indices. What
this means is that for an $S_{3}$ permutation $\sigma$, $\widehat{\lambda}_{\sigma\left(ijk\right)}^{(3;1)}$
is equal to a linear combination of $\widehat{\lambda}_{ijk}^{(3;1)}$
and $\widehat{\lambda}_{ijk}^{(3;2)}$. The same is true for $\widehat{\lambda}_{\sigma\left(ijk\right)}^{(3;2)}$.
\item Two of the $\widehat{\lambda}_{ijkl}^{(15;a)}$ parameters are symmetric
both for an exchange of the $\left(i,j\right)$ indices as well as
$\left(k,l\right)$; the remaining two $\widehat{\lambda}_{ijkl}^{(15;a)}$
are anti-symmetric if we exchange the $\left(i,j\right)$ or $\left(k,l\right)$
indices.
\end{itemize}

\subsection*{\label{sec:AppendixB}B -- Fine tunings}

Some parameters which multiply gauge invariant products of $G_{SM}$
representations have relations among themselves in an $SU(5)$-symmetric
theory. For example, in minimal $SU(5)$ it is well known that $y_{\ell}=y_{d}^{T}$
simply because $\overline{\mathbf{5}}_{F,i}\mathbf{10}_{F,j}\mathbf{5}^{*}$
decomposes as $\left(d_{i}^{c}Q_{j}+L_{i}e_{j}^{c}\right)H^{*}$ plus
interactions involving the scalar colour triplet, which is heavy.
There is no need to consider here the overall normalization factor
in the contraction of the $SU(5)$ representations but, on the other
hand, the precise way of contracting the $G_{SM}$ representations
is important therefore let us briefly state how $SU(2)$ and $SU(3)$
indices are contracted.

Take first $SU(2)$. In the case of two doublets, we assume $\mathbf{2}\cdot\mathbf{2}'\equiv\epsilon_{ab}\mathbf{2}_{a}\mathbf{2}'_{b}$
where $\epsilon$ is the Levi-Civita tensor (primes are used to distinguish
fields transforming in the same way). If there is a triplet, one can
picture its three components $\mathbf{3}_{a}$ (their electric charges
depend on the hypercharge) as forming a matrix 
\begin{align}
\left[\mathbf{3}\right]_{bc} & \equiv\left(\begin{array}{cc}
\mathbf{3}_{3} & -\frac{1}{\sqrt{2}}\mathbf{3}_{2}\\
-\frac{1}{\sqrt{2}}\mathbf{3}_{2} & \mathbf{3}_{1}
\end{array}\right)_{bc}\,,
\end{align}
in which case $\mathbf{2}\cdot\mathbf{2}'\cdot\mathbf{3}\equiv\mathbf{2}_{a}\mathbf{2}'_{b}\left[\mathbf{3}\right]_{ab}$
and $\mathbf{2}^{*}\cdot\mathbf{2}'^{*}\cdot\mathbf{3}\equiv\epsilon_{ac}\epsilon_{bd}\left(\mathbf{2}^{*}\right)_{c}\left(\mathbf{2}'^{*}\right)_{d}\left[\mathbf{3}\right]_{ab}$.

For colour indices, $\mathbf{3}\cdot\mathbf{3}'\cdot\mathbf{3}''\equiv\epsilon_{abc}\mathbf{3}_{a}\mathbf{3}'_{b}\mathbf{3}''_{c}$.
Expressions with sextets are more easily expressed if we write the
six components $\mathbf{6}_{a}$ as a symmetric matrix:
\begin{align}
\left[\mathbf{6}\right]_{bc} & \equiv\left(\begin{array}{ccc}
\mathbf{6}_{1} & \frac{1}{\sqrt{2}}\mathbf{6}_{2} & \frac{1}{\sqrt{2}}\mathbf{6}_{3}\\
\frac{1}{\sqrt{2}}\mathbf{6}_{2} & \mathbf{6}_{4} & \frac{1}{\sqrt{2}}\mathbf{6}_{5}\\
\frac{1}{\sqrt{2}}\mathbf{6}_{3} & \frac{1}{\sqrt{2}}\mathbf{6}_{5} & \mathbf{6}_{6}
\end{array}\right)_{bc}\,.
\end{align}
With this notation, $\mathbf{3}^{*}\cdot{\mathbf{3}'}^{*}\cdot\mathbf{6}\equiv\left(\mathbf{3}^{*}\right)_{a}\left({\mathbf{3}'}^{*}\right)_{b}\left[\mathbf{6}\right]_{ab}$
and $\mathbf{6}\cdot\mathbf{6}'\cdot\mathbf{6}''\equiv\epsilon_{abc}\epsilon_{def}\left[\mathbf{6}\right]_{ad}\left[\mathbf{6}'\right]_{be}\left[\mathbf{6}''\right]_{cf}$.
Finally, for both $SU(2)$ and $SU(3)$ indices, $\boldsymbol{R}\cdot\boldsymbol{R}^{*}\equiv\boldsymbol{R}_{a}\cdot\left(\boldsymbol{R}^{*}\right)_{a}$
for any representation $\boldsymbol{R}$.

Having said this, we can expand some important $SU(5)$ interactions
and express them in terms of the $G_{SM}$ fields:\footnote{These Georgi-Jarlskog-like factors can be computed in a systematic
way with the \texttt{SubgroupEmbeddingCoefficients} function of the
Susyno program \cite{Fonseca:2011sy}. The user needs to interpret/adapt
the output with care, since the program assumes a sign and normalization
convention for the contraction of $SU(5)$, $SU(2)$ and colour indices
which very likely differs from the one preferred/intended by the user.}
\begin{align}
\overline{\mathbf{5}}_{F,i}\boldsymbol{10}_{F,j}\mathbf{5}^{*}\propto\,\,\, & \left(d_{i}^{c}Q_{j}+L_{i}e_{j}^{c}\right)H'^{*}+\left(\textrm{heavy}\right)\,,\label{eq:25}\\
\overline{\mathbf{5}}_{F,i}\boldsymbol{10}_{F,j}\mathbf{45}^{*}\propto\,\,\, & \left(d_{i}^{c}Q_{j}-3L_{i}e_{j}^{c}\right)H''^{*}+\left(\textrm{heavy}\right)\,,\\
\boldsymbol{10}_{F,i}\boldsymbol{10}_{F,j}\mathbf{5}\propto\,\,\, & \left(u_{i}^{c}Q_{j}+Q_{i}u_{j}^{c}\right)H'+\left(\textrm{heavy}\right)\,,\\
\boldsymbol{10}_{F,i}\boldsymbol{10}_{F,j}\mathbf{45}\propto\,\,\, & \left(u_{i}^{c}Q_{j}-Q_{i}u_{j}^{c}\right)H''+\left(\textrm{heavy}\right)\,,\label{eq:28}\\
\overline{\mathbf{5}}_{F,i}\overline{\mathbf{5}}_{F,j}\mathbf{15}\propto\,\,\, & \left(d_{i}^{c}L_{j}+L_{i}d_{j}^{c}\right)\boldsymbol{T}-\sqrt{2}d_{i}^{c}d_{j}^{c}\boldsymbol{Z}+\left(\textrm{heavy}\right)\,,\label{eq:29}\\
\boldsymbol{10}_{F,i}\boldsymbol{10}_{F,j}\mathbf{50}\propto\,\,\, & u_{i}^{c}u_{j}^{c}\boldsymbol{Y}+Q_{i}Q_{j}\boldsymbol{X}-e_{i}^{c}e_{j}^{c}\boldsymbol{R}+\left(\textrm{heavy}\right)\,,\label{eq:30}\\
\mathbf{15}_{i}\mathbf{15}_{j}\mathbf{50}\propto\,\,\, & \boldsymbol{T}_{i}\boldsymbol{T}_{j}\boldsymbol{X}-\boldsymbol{Z}_{i}\boldsymbol{Z}_{j}\boldsymbol{Y}+\left(\textrm{heavy}\right)\,,\label{eq:31}\\
\left[\mathbf{15}_{i}\mathbf{15}_{j}\mathbf{50}\left\langle \mathbf{24}\right\rangle \right]_{1}\propto\,\,\, & \boldsymbol{T}_{i}\boldsymbol{T}_{j}\boldsymbol{X}+4\boldsymbol{Z}_{i}\boldsymbol{Z}_{j}\boldsymbol{Y}+\left(\textrm{heavy}\right)\,,\label{eq:32}\\
\mathbf{15}_{i}\mathbf{15}_{j}\mathbf{50}\left\langle \mathbf{75}\right\rangle \propto\,\,\, & \boldsymbol{T}_{i}\boldsymbol{T}_{j}\boldsymbol{X}+\boldsymbol{Z}_{i}\boldsymbol{Z}_{j}\boldsymbol{Y}+\left(\textrm{heavy}\right)\,,\label{eq:33}\\
\mathbf{15}_{i}\mathbf{50}^{*}\mathbf{50}^{*}\propto\,\,\, & \boldsymbol{Z}_{i}\left(\boldsymbol{R}^{*}\boldsymbol{Y}^{*}+\boldsymbol{Y}^{*}\boldsymbol{R}^{*}\right)-\boldsymbol{Z}_{i}\boldsymbol{X}^{*}\boldsymbol{X}^{*}+\left(\textrm{heavy}\right)\,,\label{eq:34}\\
\mathbf{15}_{i}\mathbf{50}^{*}\mathbf{50}^{*}\left\langle \mathbf{24}\right\rangle \propto\,\,\, & \boldsymbol{Z}_{i}\left(\boldsymbol{R}^{*}\boldsymbol{Y}^{*}+\boldsymbol{Y}^{*}\boldsymbol{R}^{*}\right)-\boldsymbol{Z}_{i}\boldsymbol{X}^{*}\boldsymbol{X}^{*}+\left(\textrm{heavy}\right)\,,\\
\mathbf{15}_{i}\mathbf{50}^{*}\mathbf{50}^{*}\left\langle \mathbf{75}\right\rangle \propto\,\,\, & 2\boldsymbol{Z}_{i}\left(\boldsymbol{R}^{*}\boldsymbol{Y}^{*}+\boldsymbol{Y}^{*}\boldsymbol{R}^{*}\right)+\boldsymbol{Z}_{i}\boldsymbol{X}^{*}\boldsymbol{X}^{*}+\left(\textrm{heavy}\right)\,.\label{eq:36}
\end{align}
(The physical Higgs boson $H$ is assumed to be a combination of the
doublets $H'$ and $H''$ contained in the $\mathbf{5}$ and $\mathbf{45}$,
respectively.) Interactions involving heavy $G_{SM}$ representations
are not shown.

Equations \eqref{eq:25}--\eqref{eq:28} are important for obtaining
realistic SM fermions masses; the Yukawa interaction in \eqref{eq:29}
participates in both our models, while \eqref{eq:30} is only present
in model T-II. This latter model contains scalar trilinear interactions
$\mathbf{15}\cdot\mathbf{15}\cdot\mathbf{50}$ and $\mathbf{15}\cdot\mathbf{50}^{*}\cdot\mathbf{50}^{*}$
as well, which lead to both $0\nu\beta\beta$ and neutron--antineutron
oscillations. To suppress the latter process one can tune the trilinear
coupling of $\mathbf{15}\cdot\mathbf{15}\cdot\mathbf{50}$ with the
quartic coupling of $\mathbf{15}\cdot\mathbf{15}\cdot\mathbf{50}\cdot\left\langle \mathbf{24}\right\rangle $
given that these two field contractions have different relative group-theoretical
coefficients for the $\boldsymbol{T}\boldsymbol{T}\boldsymbol{X}$
and $\boldsymbol{Z}\boldsymbol{Z}\boldsymbol{Y}$ interactions. This
is shown in equations \eqref{eq:31} and \eqref{eq:32}. On the other
hand, $\mathbf{15}\cdot\mathbf{50}^{*}\cdot\mathbf{50}^{*}$ and $\mathbf{15}\cdot\mathbf{50}^{*}\cdot\mathbf{50}^{*}\cdot\left\langle \mathbf{24}\right\rangle $
do share the same factors, so one must use instead the VEV of the
$\mathbf{75}$ to suppress neutron--antineutron oscillations (see
equations \eqref{eq:34}--\eqref{eq:36}).\\

In both models presented in this paper, one must also make sure that
it is possible, by a suitable tuning of parameters, to make the various
$G_{SM}$ representations light (i.e., significantly smaller than
the GUT scale). We shall illustrate this here only for $\boldsymbol{T}\in\mathbf{15}$
although we have checked that the same is possible for all particles
in both models. Consider then for simplicity a single copy of the
$\mathbf{15}$, which contains $\boldsymbol{T}\equiv\left(\mathbf{3},\mathbf{2},\frac{1}{6}\right)$,
$\boldsymbol{Z}\equiv\left(\mathbf{6},\mathbf{1},-\frac{2}{3}\right)$
and $\Delta\equiv S_{\mathbf{1},\mathbf{3},1}$. Once the $\mathbf{24}$
acquires a VEV, the trilinear coupling $\mathbf{15}\cdot\mathbf{\mathbf{15}}^{*}\cdot\mathbf{24}$
will contribute differently to the masses of these three $G_{SM}$
representations:
\begin{align}
\mathbf{15}\cdot\mathbf{\mathbf{15}}^{*}\cdot\left\langle \mathbf{24}\right\rangle  & \propto\,\,\,4\boldsymbol{T}\boldsymbol{T}^{*}-\boldsymbol{Z}\boldsymbol{Z}^{*}-6\Delta\Delta^{*}\,.
\end{align}
As far as quartic couplings are concerned, ignoring the scalar $\mathbf{75}$
representation (which may also acquire a VEV), one must consider three
independent quartic couplings $\widehat{\lambda}^{(5;a)}\left[\mathbf{15}\cdot\mathbf{15}^{*}\cdot\mathbf{24}\cdot\mathbf{24}\right]_{a=1,2,3}$
(confer with equation \eqref{eq:19}). It turns out that $\left[\mathbf{15}\cdot\mathbf{15}^{*}\cdot\left\langle \mathbf{24}\right\rangle \cdot\left\langle \mathbf{24}\right\rangle \right]_{a=1,2,3}\propto\,\,\,c_{a,1}\boldsymbol{T}\boldsymbol{T}^{*}+c_{a,2}\boldsymbol{Z}\boldsymbol{Z}^{*}+c_{a,3}\Delta\Delta^{*}$
with the vectors $\left(c_{a,1},c_{a,2},c_{a,3}\right)^{T}$ for $a=1,2,3$
being linearly independent so, considering that linear combinations
of the three gauge invariant contractions $\left[\mathbf{15}\cdot\mathbf{15}^{*}\cdot\mathbf{24}\cdot\mathbf{24}\right]_{a}$
are obviously gauge invariant as well, we may go ahead and define
them such that 
\begin{align}
\left[\mathbf{15}\cdot\mathbf{15}^{*}\cdot\left\langle \mathbf{24}\right\rangle \cdot\left\langle \mathbf{24}\right\rangle \right]_{1} & \propto\,\,\,\boldsymbol{T}\boldsymbol{T}^{*}\,,\\
\left[\mathbf{15}\cdot\mathbf{15}^{*}\cdot\left\langle \mathbf{24}\right\rangle \cdot\left\langle \mathbf{24}\right\rangle \right]_{2} & \propto\,\,\,\boldsymbol{Z}\boldsymbol{Z}^{*}\,,\\
\left[\mathbf{15}\cdot\mathbf{15}^{*}\cdot\left\langle \mathbf{24}\right\rangle \cdot\left\langle \mathbf{24}\right\rangle \right]_{3} & \propto\,\,\,\Delta\Delta^{*}\,.
\end{align}
Then, at the $SU(5)$-breaking scale,
\begin{align}
m_{T}^{2} & =m_{\mathbf{15}}^{2}+4\widehat{h}\left\langle \mathbf{24}\right\rangle +\widehat{\lambda}^{(5;1)}\,,\\
m_{\boldsymbol{Z}}^{2} & =m_{\mathbf{15}}^{2}-\widehat{h}\left\langle \mathbf{24}\right\rangle +\widehat{\lambda}^{(5;2)}\,,\\
m_{S_{\mathbf{1},\mathbf{3},1}}^{2} & =m_{\mathbf{15}}^{2}-6\widehat{h}\left\langle \mathbf{24}\right\rangle +\widehat{\lambda}^{(5;3)}\,,
\end{align}
where $m_{\mathbf{15}}^{2}$ is the mass term of the $\mathbf{15}$
and $\widehat{h}$ is the (properly normalized) trilinear coupling
of $\mathbf{15}\cdot\mathbf{\mathbf{15}}^{*}\cdot\mathbf{24}$ (corresponding
to what we called $\widehat{h}^{(10)}$ in equations \eqref{eq:5}
and $\widehat{h}_{ij}^{(12)}$ in \eqref{eq:18}). As such, it is
clear that with an appropriate tuning of parameters, it is possible
to make $\boldsymbol{T}$ and $\boldsymbol{Z}$ light while keeping
$\Delta$ heavy.

\subsection*{\label{sec:AppendixC}C -- Decomposition table of $SU(5)$ representations}

For reference, we provide here a table with the decomposition of $SU(5)$
representations into those of the SM group (this data coincides with
the one given in \cite{Slansky:1981yr}).

\begin{center}
\begin{table}[tbph]
\begin{centering}
\begin{tabular}{cc}
\hline
 Rep. & Decomposition\tabularnewline
\hline
\textbf{5} & $\left(\mathbf{3},\mathbf{1},-\frac{1}{3}\right)+\left(\mathbf{1},\mathbf{2},\frac{1}{2}\right)$\tabularnewline
\textbf{10} & $\left(\overline{\mathbf{3}},\mathbf{1},-\frac{2}{3}\right)+\left(\mathbf{3},\mathbf{2},\frac{1}{6}\right)+\left(\mathbf{1},\mathbf{1},1\right)$\tabularnewline
\textbf{15} & $\left(\mathbf{6},\mathbf{1},-\frac{2}{3}\right)+\left(\mathbf{3},\mathbf{2},\frac{1}{6}\right)+\left(\mathbf{1},\mathbf{3},1\right)$\tabularnewline
\textbf{24} & $\left(\mathbf{3},\mathbf{2},-\frac{5}{6}\right)+\left(\mathbf{8},\mathbf{1},0\right)+\left(\mathbf{1},\mathbf{3},0\right)+\left(\mathbf{1},\mathbf{1},0\right)+\left(\overline{\mathbf{3}},\mathbf{2},\frac{5}{6}\right)$\tabularnewline
\textbf{35} & $\left(\overline{\mathbf{10}},\mathbf{1},1\right)+\left(\overline{\mathbf{6}},\mathbf{2},\frac{1}{6}\right)+\left(\overline{\mathbf{3}},\mathbf{3},-\frac{2}{3}\right)+\left(\mathbf{1},\mathbf{4},-\frac{3}{2}\right)$\tabularnewline
\textbf{40} & $\left(\mathbf{8},\mathbf{1},1\right)+\left(\overline{\mathbf{6}},\mathbf{2},\frac{1}{6}\right)+\left(\mathbf{3},\mathbf{2},\frac{1}{6}\right)+\left(\overline{\mathbf{3}},\mathbf{3},-\frac{2}{3}\right)+\left(\overline{\mathbf{3}},\mathbf{1},-\frac{2}{3}\right)+\left(\mathbf{1},\mathbf{2},-\frac{3}{2}\right)$\tabularnewline
\textbf{45} & $\left(\overline{\mathbf{3}},\mathbf{1},\frac{4}{3}\right),\left(\mathbf{8},\mathbf{2},\frac{1}{2}\right),\left(\mathbf{1},\mathbf{2},\frac{1}{2}\right),\left(\overline{\mathbf{6}},\mathbf{1},-\frac{1}{3}\right),\left(\mathbf{3},\mathbf{3},-\frac{1}{3}\right),\left(\mathbf{3},\mathbf{1},-\frac{1}{3}\right),\left(\overline{\mathbf{3}},\mathbf{2},-\frac{7}{6}\right)$\tabularnewline
\textbf{50} & $\left(\mathbf{6},\mathbf{1},\frac{4}{3}\right)+\left(\mathbf{8},\mathbf{2},\frac{1}{2}\right)+\left(\overline{\mathbf{6}},\mathbf{3},-\frac{1}{3}\right)+\left(\mathbf{3},\mathbf{1},-\frac{1}{3}\right)+\left(\overline{\mathbf{3}},\mathbf{2},-\frac{7}{6}\right)+\left(\mathbf{1},\mathbf{1},-2\right)$ \tabularnewline
\textbf{70} & $\left(\mathbf{6},\mathbf{2},-\frac{7}{6}\right)+\left(\mathbf{15},\mathbf{1},-\frac{1}{3}\right)+\left(\mathbf{3},\mathbf{3},-\frac{1}{3}\right)+\left(\mathbf{3},\mathbf{1},-\frac{1}{3}\right)+\left(\mathbf{8},\mathbf{2},\frac{1}{2}\right)+\left(\mathbf{1},\mathbf{4},\frac{1}{2}\right)+\left(\mathbf{1},\mathbf{2},\frac{1}{2}\right)+\left(\overline{\mathbf{3}},\mathbf{3},\frac{4}{3}\right)$\tabularnewline
\textbf{70'} & $\left(\overline{\mathbf{15'}},\mathbf{1},\frac{4}{3}\right)+\left(\overline{\mathbf{10}},\mathbf{2},\frac{1}{2}\right)+\left(\overline{\mathbf{6}},\mathbf{3},-\frac{1}{3}\right)+\left(\overline{\mathbf{3}},\mathbf{4},-\frac{7}{6}\right)+\left(\mathbf{1},\mathbf{5},-2\right)$ \tabularnewline
\textbf{75} & $\left(\overline{\mathbf{3}},\mathbf{1},-\frac{5}{3}\right)+\left(\overline{\mathbf{6}},\mathbf{2},-\frac{5}{6}\right)+\left(\mathbf{3},\mathbf{2},-\frac{5}{6}\right)+\left(\mathbf{8},\mathbf{3},0\right)+\left(\mathbf{8},\mathbf{1},0\right)+\left(\mathbf{1},\mathbf{1},0\right)+\left(\mathbf{6},\mathbf{2},\frac{5}{6}\right)+\left(\overline{\mathbf{3}},\mathbf{2},\frac{5}{6}\right)+\left(\mathbf{3},\mathbf{1},\frac{5}{3}\right)$\tabularnewline
\hline
\end{tabular}
\par\end{centering}

\protect\caption{\label{tab:decopm}$SU(3)\times SU(2)\times U(1)$ decomposition of some $SU(5)$ representations.
As usual, the correctly normalized hypercharge $y_{N}$ is obtained
by multiplying the values of $y$ shown here by $\sqrt{\frac{3}{5}}$.}
\end{table}

\par\end{center}

\end{document}